\documentclass[lettersize,journal]{IEEEtran}
\usepackage{amsmath,amsfonts}
\usepackage{algorithmic}
\usepackage{algorithm}
\usepackage{array}
\usepackage[caption=false,font=normalsize,labelfont=sf,textfont=sf]{subfig}
\usepackage{textcomp}
\usepackage{stfloats}
\usepackage{url}
\usepackage{verbatim}
\usepackage{graphicx}
\usepackage{cite}
\usepackage{stfloats}
\usepackage{cuted} % 加载cuted宏包
\usepackage{amsmath,amsfonts}
\usepackage{algorithmic}
\usepackage{algorithm}
\usepackage{array}
\usepackage[caption=false,font=normalsize,labelfont=sf,textfont=sf]{subfig}
\usepackage{textcomp}
\usepackage{stfloats}
\usepackage{url}
\usepackage{verbatim}
\usepackage{graphicx}
\usepackage{cite}
\usepackage{color}
\usepackage{graphicx}
\usepackage{makecell} 
\usepackage{multicol}
\usepackage{booktabs}
\usepackage{hyperref}
\usepackage{stfloats}
\begin{document}

\title{ Multimodal Visual Image Based User Association and Beamforming Using Graph Neural Networks}

\author{Yinghan Li,~\IEEEmembership{Graduate Student Member,~IEEE}, Yiming Liu,~\IEEEmembership{Graduate Student Member,~IEEE} \\ and Wei Yu,~\IEEEmembership{Fellow,~IEEE}% <-this % stops a space
\thanks{The authors are with The Edward S. Rogers Sr. Department of Electrical and Computer Engineering, University of Toronto, Toronto, ON M5S 3G4, Canada. E-mail: \{yinghan.li, eceym.liu\}@mail.utoronto.ca; weiyu@ece.utoronto.ca.  This work was supported by
Natural Sciences and Engineering Research Council via the Canada Research Chairs program.}% <-this % stops a space
%\thanks{Manuscript received April 19, 2021; revised August 16, 2021.}
}



\maketitle

\thispagestyle{empty}

\begin{abstract}

This paper proposes an approach that leverages multimodal data by integrating
visual images with radio frequency (RF) pilots to optimize user association and
beamforming in a downlink wireless cellular network under a max-min fairness
criterion.  Traditional methods typically optimize wireless system parameters
based on channel state information (CSI). However, obtaining accurate CSI requires
extensive pilot transmissions, which lead to increased overhead and latency. 
Moreover, the optimization of user association and beamforming is a discrete
and non-convex optimization problem, which is challenging to solve analytically.  
In this paper, we propose to incorporate visual camera data in addition to the 
RF pilots to perform the joint optimization of user association and beamforming. 
The visual image data help enhance channel awareness, thereby reducing the dependency on extensive 
pilot transmissions for system optimization. We employ a learning-based approach 
based on using first a detection neural network that estimates user locations 
from images, and subsequently two graph neural networks (GNNs) that extract 
features for system optimization based on the location information and the 
received pilots, respectively. Then, a multimodal GNN is constructed to 
integrate the features for the joint optimization user association and beamforming.  
Simulation results
demonstrate that the proposed method achieves superior performance, while 
having low computational complexity and being interpretable and generalizable, 
making it an effective solution as compared to traditional methods based only on RF pilots.

\end{abstract}


\section{Introduction}
In wireless communication, deploying multiple access points (APs) is an efficient approach to enhance network coverage, capacity, and reliability in densely populated urban areas with numerous building obstacles.
In multi-AP systems, finding efficient user association, power allocation, and beamforming strategies based on the channel state information (CSI) is the central task for optimizing system performance \cite{Xu21survey}. However, obtaining accurate CSI between multiple access points (APs) and user equipments (UEs) solely from RF pilots is a challenging and resource-intensive task. Additionally, realizing coordinated optimization among multiple transmitters is a nontrivial problem. The main objective of this paper is to enhance channel awareness by integrating insights from diverse data sources, enabling more efficient data-driven solutions for complex user association and beamforming optimization.

This paper proposes a method that leverages both visual data from images and radio frequency (RF) pilots to address the downlink user association and beamforming problem, with an aim to maximize the minimum achievable rate across all users and ensure fairness. Based on the pilots and images, a central server assigns each user to an AP and designs the downlink beamformers for all the target users for the AP under the power constraints to optimize certain system-level objective. 
In densely deployed cellular networks, user mobility across the cells leads to frequent fluctuations in CSI, which requires accurate and consistent re-estimation in order to maintain effective network performance. In this case, conventional methods, which rely solely on pilots to achieve highly accurate channel estimation \cite{Coleri}, would result in substantial overhead and latency.
Visual data offers the advantage of providing complementary real-time insights into the propagation environment and channel characteristics without consuming spectral resources. This motivates the approach proposed in this paper for integrating visual images with pilot signals to enhance channel awareness, thereby reducing the reliance on extensive pilot transmissions for accurate CSI estimation.

Multimodal data based user association and beamforming optimization is complex due to the challenge of aligning and integrating different types of data.
In this paper, we propose to use neural networks to detect and to extract user locations from the visual images, and to perform system optimization based on the channel characteristics information implicitly contained in the user locations.
The proposed method bypasses explicit CSI estimation from the two data sources, and employs two graph neural networks (GNNs) \cite{hamilton} to extract essential features to perform user association and beamforming. The first GNN extracts features from the locations estimated from the images, and another GNN extracts features from RF pilots. 
Specifically, the two GNNs process their inputs independently, then refine feature representations in successive neural network layers to extract the specific channel characteristics needed for system-level optimization based on their respective input data types. 
A third GNN is then introduced to integrate the updated features extracted by the two GNNs, and to further refine the combined information for improved performance and more efficient decision-making.

This paper adopts GNNs %to extract essential features for optimizing system parameters based on multimodal data.  GNNs are chosen 
for their ability to capture underlying graph-based relationships, which makes them particularly effective for optimization tasks that involve parameters with complex interdependencies. In this paper, each node in the GNN represents a UE and is associated with a user association decision and an assigned beamforming vector.  The graphical structure enables the GNN to determine the optimal allocation strategy for each user, while leveraging the dependencies between users, such as the inter-user interference which is a function of their spatial locations. 
In addition, the GNN-based framework has generalization capability that enables a single model to adapt seamlessly across varying network scenarios with different number of UEs.  The capacity to handle complex interdependencies and the capability for generalization make GNNs a robust and adaptable framework for solving user association and beamforming optimization problems in cellular networks.
 
\subsection{Related Works}
The user association problem has been considered extensively in the literature\cite{surveyliu}.
 The simplest way to determine the user associations is based on the signal-to-noise ratio (SNR), where each UE connects to the AP that provides the strongest SNR \cite{goldsmith2005wireless}. 
The limitation of the max-SNR strategy is that it often results in unfairness due to uneven loading at the APs when the UE distribution is uneven, leading to poor service for users connected to congested APs.
In many works, it has been demonstrated that strategically optimizing system configurations can achieve better performance than the {max-SNR} strategy. The optimized strategy is designed based on various objective functions, such as enhancing power efficiency \cite{Mesodia}, maximizing sum rate \cite{Corroy}, and improving fairness \cite{Sun2015}. The works in the area of power efficiency mostly focus on minimizing the total power consumption while maintaining the required quality of service (QoS) for users\cite{Qian,Qian2,Zhou,Yates95,Rashid97}.
The works in the area of rate maximization primarily focus on developing strategies to optimize resource allocation and user association to ensure that the combined or weighted data rates of all users are maximized across the network\cite{Corroy,Sanjabi,Cui2,Zhao18}.   

This paper focuses on a fairness-oriented approach to solve the user association and resource allocation problem. Research in this area typically seeks fairness by employing objective functions such as proportional fairness and max-min fairness.
%The user association problem, with fairness as the objective function, has been proven to be challenging due to its nonconvex nature and the involvement of discrete constraints.
In this realm, traditional model based methods\cite{Zhang17,Sun2015,Liu15} address the user association problem by simplifying the problem into a more manageable form — either by optimizing an upper bound or by relaxing the binary constraints into continuous ones.  
In contrast, more recent works have proposed data-driven methods, for example \cite{Jang22}, which utilizes deep reinforcement learning (DRL) to tackle the user association problem.  
However, these works primarily focus on single-antenna scenarios \cite{Sun2015,Zhang17,Liu15,Jia16,Jang22}, limiting their applicability in more complex multi-antenna environments, where the use of beamforming may require a very different association strategy as compared to single-antenna setups.

The studies \cite{kaiming,Wang22} explore analytic methods to address beamforming design, power allocation, and user association under the fairness criterion in multi-antenna scenarios.
In \cite{kaiming}, the authors propose a distributed pricing-based method that leverages a dual coordinate descent approach to optimize the system configuration. Using weighted rate maximization as the objective function, fairness is achieved by defining the weight as the reciprocal of each user’s long-term average rate.
However, long-term fairness cannot guarantee user QoS in the short term.  This brings us to the instantaneous max-min fairness criterion, which helps users with the weakest channel conditions by maximizing the minimum rate in each scheduling instance. In \cite{Sanjabi}, the instantaneous $\alpha$-proportional fairness objective function is introduced to guarantee the fairness. Although the authors suggest that setting $\alpha$ to infinity theoretically approaches max-min fairness, the method in \cite{Sanjabi} can suffer from  numerical instability when $\alpha$ is very large, making it impractical for solving problems under the max-min fairness criterion.
Optimization under the instantaneous max-min fairness criterion is considered in \cite{Wang22}. However, the authors simplify the optimization problem by ignoring the interference and by replacing the constraint that each user must be assigned to only one AP with a constraint that limits the number of users each AP can serve.
The user association and beamforming problem can also be tackled using exhaustive search, which evaluates all possible user association configurations to find the optimal one. For each configuration, the corresponding multipoint beamforming problem can be solved using the method proposed in \cite{multipoint}. However, this approach is impractical for large-scale systems due to its high complexity.
Finally, all the above methods assume high-accuracy CSI, which has to be obtained from extensive pilot signals.  
A key contribution of the proposed method in this paper is its ability to enhance system performance without extensive pilot transmissions by leveraging multimodal data.

One of the key differentiating features of this paper, compared to conventional approaches, is the utilization of visual data to enhance communication efficiency\cite{Tian21}. Prior studies have demonstrated the potential of visual information for solving key tasks in wireless communication, such as channel estimation\cite{Xu21}, handover \cite{Charan21} and beamforming \cite{Jiang22,Xu23,Lin24,Charan22,Weihua23}. In the context of beamforming, most prior works, such as \cite{Jiang22,Xu23}, focus on optimizing beamforming by maximizing signal-to-noise ratio (SNR) based on the image data. However, these approaches simplify the problem by omitting interference considerations, so they can lead to suboptimal solutions in realistic multi-user environments. For example, \cite{Charan22,Lin24} propose using multimodal data, including image combined with accurate channel information and the user location information for beamforming, but they do not fully address the interference among the users. In \cite{Weihua23}, interference is considered only for power allocation optimization, while the beamforming design remains focused on maximizing SNR, which may not be ideal in multi-user scenarios. Moreover, all these methods rely on supervised learning, which requires optimal or suboptimal labels for large training datasets—an often unrealistic assumption in complex, real-world wireless networks. 
Additionally, none of these works consider the user association optimization without CSI.

\subsection{Contributions}
This paper considers the joint user association and beamforming optimization problem under a max-min fairness objective. The main contributions of this paper are summarized as follows\footnote{{The codes and dataset for this paper are available at: \url{https://github.com/Leeyyhh/Multimodal-Based-User-Association-and-Beamforming}}.}.

%In this paper, we propose a novel approach that integrates visual data from cameras with RF signals to solve the joint user association and beamforming problem in downlink MISO systems. The contributions are summarized as follows:
\subsubsection{Multimodal Data Integration for Enhanced CSI Awareness}
This paper introduces a multimodal approach that combines visual images and RF pilots to optimize user association and beamforming. In mobile user scenarios, accurate channel information is usually a prerequisite for ensuring optimal system performance. However, relying solely on pilot signals for CSI estimation leads to significant overhead and latency. To solve this problem, this paper integrates visual data with RF pilots to provide real-time channel insights without consuming additional spectral resources. Specifically, our approach employs two GNNs to independently learn critical features for optimization from RF data and from location information derived from the images. 
A third GNN then integrates these features to optimize user association and beamforming, in effect combining insights from both data sources to enable comprehensive decision-making. 
The simulation results demonstrate significant performance gains of the proposed multimodal data based method over pilot-based benchmark methods. %, showcasing its effectiveness in enhancing system performance.
%We propose a flexible framework capable of utilizing multimodal data to optimize user association and beamforming. By integrating both visual and RF pilot data, the framework enhances the wireless propagation environment awareness, leading to more accurate and efficient decision-making in the proposed problem. Moreover, when either RF or image data is unavailable, the framework can adapt by optimizing based on the remaining data, ensuring flexibility and reliable operation across various conditions.
\subsubsection{Detection Network for Location Estimation}
We propose a location detection stage to map image data captured by cameras at APs to the estimated user locations. The network includes a well-trained location detection model that identifies and localizes targets within the images, a filtering block that refines the detection results, and a calibration matrix that maps pixel coordinates from the images to the real-world coordinates. Additionally, we provide a detailed analysis of location estimation errors to demonstrate the effectiveness of the detection network, and to show that it is a reliable tool for enhancing environmental awareness. 

\subsubsection{GNN-based Neural Network for User Association and Beamforming}
This paper proposes a GNN-based neural network to optimize user association and beamforming based on multimodal data, i.e., the estimated user locations and the pilots transmitted by the users. The user association and beamforming problem is challenging due to the need for coordinated decision-making across APs, where each user’s association and beamforming configuration both affects and is affected by others.  In addition, dynamic network scenarios require adaptability to varying user distributions. In the proposed GNN, each user is modelled as a node. This GNN-based framework excels at capturing complex interdependencies among the nodes, and has strong generalization ability, enabling it to perform effectively across scenarios with varying numbers of users.

\subsubsection{Unsupervised Learning with a Modified Straight-Through Estimator (STE) for Binary User Association}
In this paper, we utilize unsupervised learning to avoid the complexity of obtaining labels for non-convex optimization problem with discrete user associations constraints.
However, directly quantizing the outputs from the proposed GNN into the required binary user associations during training disrupts gradient flow from the loss function to the neural network coefficients.
To maintain accurate and stable gradient backpropagation, the STE is employed to enable gradients to flow through despite the non-differentiability.
Furthermore, since user associations act as coefficients for beamforming vectors, setting some user associations to 0 in the early stages of training can negatively impact the optimization process. To prevent this, we propose a modified STE that avoids the premature exclusion of user-AP pairs and ensures continuous updates to both user associations and beamforming vectors throughout the training process.

\subsection{Paper Organization and Notations}
The rest of this paper is organized as follows.
Section II introduces the problem formulation.
Section III describes the proposed image-based location estimation neural network.
Section IV outlines the proposed GNN framework for solving the user association and beamforming problem. 
Sections V and VI present simulation results and interpret the learned solutions from the GNNs.
Section VII concludes the paper.

The notations used in this paper are as follows: Scalars are denoted by lowercase letters. Column vectors are denoted by bold lowercase letters. Matrices are denoted by bold uppercase letters. The $i$-th element of the vector $\mathbf{a}$ is denoted by $\mathbf{a}_i$. The transpose and Hermitian transpose of matrices are represented by $(\cdot)^{\top}$ and $(\cdot)^{\mathrm{H}}$, respectively. The complex Gaussian distribution is represented by $\mathcal{CN}(\cdot, \cdot)$, and expectation is denoted by $\mathbb{E}[\cdot]$. Additionally, $(\cdot)^\text{I}$, $(\cdot)^\text{P}$, and $(\cdot)^\text{C}$ are used to indicate components associated with image-based, pilot-based, and combined data-based optimization, respectively. Additionally, $(\cdot)^\text{I}$, $(\cdot)^\text{P}$, and $(\cdot)^\text{C}$ are used to indicate components associated with image-based, pilot-based, and combined multimodal data-based optimization, respectively.

\section{Problem Formulation}

\subsection{System Model and Problem Formulation}

Consider a downlink wireless multi-user multiple-input single-output (MU-MISO) system where $L$ APs intend to communicate to $K$ mobile UEs. Each AP is equipped with a camera and a planar antenna array of $M$ antennas, while each UE is equipped with a single antenna. 
Let $ i $ and $ k $ be the indices of UEs, where $ i, k \in \{1,2, \ldots, K\} $, and let $l$ be the index of APs, where $l \in\{1,2, \ldots, L\}$. The $l$-th AP is represented by $\mathcal{B}_l$, and the $k$-th mobile UE is represented by the $\mathcal{U}_k$. Let $ a_{kl} $ denote whether or not $ \mathcal{U}_k $ is associated with AP $ \mathcal{B}_l $, where $ a_{kl} $ should satisfy
\begin{equation}
\label{eq:optimization_d}
a_{kl} \in \{0,1\}.
\end{equation}

Since each UE is assumed to be uniquely associated with a single AP, we have 
\begin{equation}
\label{eq:optimization_c}
\sum_{l=1}^L a_{kl} = 1, \forall k.
\end{equation}
Let $\mathbf{v}_{kl} \in \mathbb{C}^M$ be the transmit beamforming vector from $\mathcal{B}_l$ intended for $\mathcal{U}_k$. The beamforming vectors need to satisfy a sum power constraint 
{
 \begin{equation}
 \label{eq:optimization_b}
\sum\limits_{k=1}^K||\mathbf{v}_{kl}||^2_2 \leq {p}_l, \forall l ,
 \end{equation} where \( ||\mathbf{v}_{kl}||^2_2 \) represents the transmit power allocated to \( \mathcal{U}_k \) by \( \mathcal{B}_l \), and  ${p}_l$ represents the maximum transmit power of $\mathcal{B}_l$.}
We assume that all APs share the same frequency and time resources for communication. The block-fading model is adopted in this paper, where the channel remains constant within each fading block but varies independently between different blocks. Let $\mathbf{h}_{kl} \in \mathbb{C}^M$ denote the channel vectors from $\mathcal{B}_l$ to $\mathcal{U}_k$. The set of all channel vectors is denoted by $\mathbf{h} = \left[ \mathbf{h}_{11}^\top, \mathbf{h}_{12}^\top, \ldots, \mathbf{h}_{1L}^\top, \mathbf{h}_{21}^\top, \ldots, \mathbf{h}_{KL}^\top \right]^\top$. Given $\mathbf{h}$, the received signal at $\mathcal{U}_k$ is given by
{
\begin{equation}
 r_k=\sum_{l=1}^L{a}_{k l} \mathbf{h}_{k l}^\top \mathbf{v}_{kl} s_k+\sum_{l=1}^L \sum_{i\neq k} {a}_{il} \mathbf{h}_{kl}^\top \mathbf{v}_{il} s_i+n_k,
\end{equation}}where $n_k \sim \mathcal{C N}(0, \sigma_k^2)$ is the additive Gaussian noise, and $s_k$ denotes the data symbol intended for $\mathcal{U}_k$, with $\mathbb{E}[|s_k|^2]=1$. The achievable rate of $\mathcal{U}_k$ can be expressed as
%$$ R_{k} =\sum_{l=1}^L {a}_{kl} \log \left( 1 + \dfrac{|\mathbf{h}_{kl} \mathbf{v}_{kl}|^2}{ \sum_{(i', j') \neq (k, j)} | \mathbf{h}_{i'j'} \mathbf{v}_{i'j'}|^2 + \sigma_k^2    }\right)  $$
\begin{equation}
 R_{k} = \log \left( 1 + \dfrac{\sum_{l=1}^L {a}_{kl}|\mathbf{h}_{kl}^\top \mathbf{v}_{kl}|^2}{ \sum_{l=1}^L \sum_{i\neq k} {a}_{il}| \mathbf{h}_{kl}^\top \mathbf{v}_{il}|^2 + \sigma_k^2 }\right).
\label{eq:rate} 
 \end{equation}
 
The objective of the user association and beamforming optimization in this paper is to ensure fairness among users by maximizing the minimum achievable rate across all users. 
This problem can be formulated mathematically as:
\begin{subequations}
\label{eq:min_rate_optimization}
\begin{align} 
\underset{
  (\mathbf{a}, \mathbf{v})
}{\operatorname{maximize}} \quad & \quad
  \min\limits_{ k } R_k \label{eq:minrate_a}\\
\text{subject to} \quad & \quad \eqref{eq:optimization_d},\eqref{eq:optimization_c},\eqref{eq:optimization_b},
\end{align}
\end{subequations}  
where $\mathbf{a}=[{a}_{11}, {a}_{12},\cdots,{a}_{1L}, {a}_{21},\cdots,{a}_{KL}]^\top$ is the user association vector, and $\mathbf{v}=[\mathbf{v}_{11}, \mathbf{v}_{12}, \ldots, \mathbf{v}_{1L}, \mathbf{v}_{21}, \ldots, \mathbf{v}_{KL}]^\top$ denotes the collection of all beamforming vectors. 

Accurately calculating the achievable rates for all users is a required first step for optimizing the minimum achievable rate across all users, and the channel $\mathbf{h}$ is needed for calculating the rate as formulated in \eqref{eq:rate}.
Thus, for optimizing the user rate as a function of user association and beamforming analytically, the availability of CSI is crucial. 

%\subsection{Channel Estimation via Pilot}
In this paper, we assume a time-division-duplex (TDD) system with uplink-downlink channel reciprocity, so that the CSI can be estimated through uplink pilots. During the uplink pilot transmission phase, each UE transmits $N_u$ orthogonal pilots. Let $\bar{\mathbf{s}}_k \in \mathbb{C}^{N_u}$ denote the pilot sequence transmitted from $\mathcal{U}_k$.  The pilot sequence should satisfy the power constraint $\bar{\mathbf{s}}_k^{\mathrm{H}} \bar{\mathbf{s}}_k = \bar{p} N_u$, where $\bar{p}$ denotes the transmit power of each pilot. The received pilots at $\mathcal{B}_l$ are represented by $\bar{\mathbf{Y}}_l \in \mathbb{C}^{M \times N_u}$, given by:
\begin{equation}
\bar{\mathbf{Y}}_l=\sum_{k=1}^K \mathbf{h}_{kl} \bar{\mathbf{s}}_k^{\mathrm{H}}+\bar{\mathbf{N}},
\end{equation}
where {$\bar{\mathbf{N}} \in \mathbb{C}^{M \times N_u}$} represents a noise matrix with each element independently distributed as $ \mathcal{C N}(0, \sigma_p^2)$. 

The pilot assignment for different users follows the traditional pilot transmission protocol, where different orthogonal pilots are assigned to different users, i.e., $\bar{\mathbf{s}}_k^{\mathrm{H}} \bar{\mathbf{s}}_i = 0$ if $ i \neq k $. 
The AP correlates $\bar{\mathbf{Y}}_l$ with the known pilot sequence $\bar{\mathbf{s}}_k$ to isolate the channel of $\mathcal{U}_k$:
\begin{equation}
\bar{\mathbf{r}}_{kl}=\frac{1}{\bar{p}N_u} \bar{\mathbf{Y}}_l \bar{\mathbf{s}}_k=\mathbf{h}_{kl}+ \frac{1}{\bar{p}N_u}{\bar{\mathbf{N}}}\bar{\mathbf{s}}_k,
\label{eq:pilot2}
\end{equation}
where $\bar{\mathbf{r}}_{kl} \in \mathbb{C}^{M \times 1}$ represents the estimated channel between $\mathcal{U}_k$ and $\mathcal{B}_l$. In this paper, we denote the received RF pilot from $K$ UEs to $L$ APs as $\bar{\mathbf{R}} \in \mathbb{C}^{M \times KL}$, where $\bar{\mathbf{R}} = \left[ \bar{\mathbf{r}}_{11}, \bar{\mathbf{r}}_{12}, \dots, \bar{\mathbf{r}}_{1L}, \bar{\mathbf{r}}_{21}, \dots, \bar{\mathbf{r}}_{KL} \right]$, and $\bar{\mathbf{r}}_{kl}$ is obtained from \eqref{eq:pilot2}.

%However, as observed from equation \eqref{eq:pilot2}, 
We note here that the accuracy of channel estimation is directly influenced by the number of available pilots $N_u$. Traditional optimization method typically assumes perfect CSI. Inaccuracies in channel estimation due to limited pilots can degrade overall system performance. While extensive pilot transmission can improve channel estimation accuracy, it also results in increased latency and overhead. To prevent excessive pilot transmission while enhancing channel awareness, this paper proposes to utilize visual data as a complementary source of channel information.  The images captured by cameras at each AP provide estimates of the target UE’s location, which is a primary factor influencing channel characteristics. By leveraging both RF pilots and location information derived from images, the system can more easily learn the channel characteristics, even in scenarios with limited pilots.

\subsection{Visual Data for Enhanced Channel Awareness}
In this paper, we assume that each AP is equipped with a camera to capture images, and each AP processes these images locally to estimate the locations of the UEs. The location information is then transmitted to the central server, where multiview spatial information from different APs is utilized for subsequent system optimization. We assume that the UE locations remain static within each fading block, which is consistent with the assumption of block-fading channel models.
Let $ \mathbf{I}_l \in \mathbb{R}^{S_x \times S_y \times 3} $ represent the image taken at $ \mathcal{B}_l $, where $ S_x $ and $ S_y $ represent the width and height of the image in pixels, respectively, and the third dimension corresponds to the three color channels (RGB).   Let $\hat{\mathbf{q}}_{kl} \in \mathbb{R}^{3 \times 1}$ represent the location information of $\mathcal{U}_k$ from image $\mathbf{I}_l$. If $\mathcal{U}_k$ is not detected in the image $\mathbf{I}_l$, set $\hat{\mathbf{q}}_{kl} = -\mathbf{1} $, where $ \mathbf{1} $ represents an all-ones vector. The $\hat{\mathbf{q}}_{kl}$ can be obtained from the image $ \mathbf{I}_l $ using a function $ \mathcal{G}_l $, defined as:
\begin{equation}
\{\hat{\mathbf{q}}_{kl}\}_{k=1}^K = \mathcal{G}_l(\mathbf{I}_l).
\label{eq:location_est}
\end{equation}

{Typically, large-scale fading coefficients, which reflect long-term channel conditions and can be estimated based on location information, are directly utilized for user association. However, in urban environments, buildings, roads, and other obstacles can cause persistent and significant multipath propagation and obstruct the propagation of signal paths. In such scenarios, relying solely on location information is insufficient. {Therefore, some recent studies consider multipath channel models to more accurately capture the propagation environment. By taking into account reflected paths, these approaches aim to provide a more realistic basis for user association and beamforming optimization \cite{multipath1,multipath2,multipath3}.} In this paper, multipath effects are considered as part of the large-scale fading characteristics, and a learning-based method is employed to capture the complex channel characteristics.}

{Let $N^R_{kl}$ represent the number of paths from $\mathcal{U}_k$ to $\mathcal{B}_l$. 
The relationship between the multipath components and the channel $\mathbf{h}_{kl}$ can be expressed as:  
\begin{equation}
\mathbf{h}_{kl} = \sum_{n=1}^{N^R_{kl}} \zeta(d_{nkl}) \mathbf{a}(\theta_{nkl}, \phi_{nkl})+\mathbf{n}_{kl} , \end{equation}
where $d_{nkl}$ represents the distance of the $n$-th path from $\mathcal{U}_k$ to $\mathcal{B}_l$, and $\theta_{nkl}$ and $\phi_{nkl}$ represent the elevation angle and azimuth angle of arrival of the $n$-th path from $\mathcal{U}_k$ to $\mathcal{B}_l$, respectively.  The parameter $\zeta(d_{nkl})$ is the large-scale fading coefficient associated with the respective path. The term $\mathbf{a}(\theta_{nkl}, \phi_{nkl})$ represents the antenna array response vector at $\mathcal{B}_l$ based on the angles of arrival $\theta_{nkl}$ and $\phi_{nkl}$. 
The term $\mathbf{n}_{kl} \in \mathbb{C}^{M}$ represents the channel characteristics that are not explicitly captured by the channel model.}

The distance $d_{nkl}$ and the angles of arrival $\phi_{nkl}$ and $\theta_{nkl}$ are determined by the locations of the transmitters, APs, and objects that reflect signals in the environment.
Under the assumption that the locations of the APs and most of the objects, such as buildings and roads, are fixed, the CSI can be approximately determined from the locations of the transmitters. Let $\hat{\mathbf{Q}} = [\hat{\mathbf{q}}_{11}, \hat{\mathbf{q}}_{12}, \dots, \hat{\mathbf{q}}_{KL}]$ denote the location matrix. The relationship between the location of UEs and CSI can be expressed by rewriting (10) using a deterministic function $\mathcal{M}$ as:
\begin{equation}
\mathbf{h} = \mathcal{M}(\hat{\mathbf{Q}})+\hat{\mathbf{N}} ,
\label{eq:channel_location}
\end{equation}
where $\hat{\mathbf{N}} \in \mathbb{C}^{MKL}$ is a matrix representing environmental factors that influence the channel characteristics independently of the location matrix $\hat{\mathbf{Q}}$.  

To realize the function $\mathcal{G}_l(\cdot)$ in equation \eqref{eq:location_est}, we propose to utilize a detection neural network to map the images to the UE locations. Advances in detection neural network models can now allow the location matrix $\hat{\mathbf{Q}}$ to be accurately obtained from the images\cite{srivastava2021comparative}. To further enhance the accuracy and robustness of $\hat{\mathbf{Q}}$, multiview images are utilized to increase the likelihood of detecting UEs and to provide location information from multiple perspectives, which collectively mitigate the effects of occlusions and detection errors.  In this way, the location estimate $\hat{\mathbf{Q}}$ can be made highly reliable. 

The matrix $\hat{\mathbf{Q}}$ can then be used to infer channel characteristics, because location information directly influences key parameters such as path loss, angles of arrival, and line-of-sight (LoS) conditions.
Based on the above discussion and the relationships described in \eqref{eq:location_est} and \eqref{eq:channel_location}, we see that the multiview images captured by AP cameras can provide reliable information for inferring the channel $\mathbf{h}$. This motivates the use of images as a secondary data source that can provide information complementary to the RF pilots, which can be used to support system optimization.

\subsection{Multimodal Data-Driven Optimization}

In this paper, we aim to optimize system configurations using received pilots and captured images.  Traditional systems typically involve a two-step process, where the channels are first estimated based on received data, and system variables are then optimized accordingly.
Since channel estimation primarily serves as an intermediate step for beamforming and user association optimization, we propose a data-driven approach that eliminates the need for explicit channel estimation.  The proposed method directly maps received pilots and image-derived location data to optimized user association and beamforming decisions.
Let $\mathcal{F}(\cdot)$ represent the function that maps the user location matrix $\hat{\mathbf{Q}}(t) $ and the RF pilots $\bar{\mathbf{R}}(t)$ at the $t$-th time slot to the optimized system configurations $\mathbf{a}(t)$ and $\mathbf{v}(t)$, defined as:
\begin{align} 
 (\mathbf{a}(t), \mathbf{v}(t))&= \mathcal{F}\left(\hat{\mathbf{Q}}(t) ,\bar{\mathbf{R}}(t)\right) 
\label{eq:channel_est}
\end{align}
Then the multimodal data based optimization can be formulated as 
\begin{subequations}
\label{eq:optimization}
\begin{align} 
\underset{
  (\mathbf{a}(t), \mathbf{v}(t))
}{\operatorname{maximize}} \quad & \quad
  \min\limits_{ k } R_k(t) \label{eq:optimization_a}\\
\text{subject to} \quad & \quad \eqref{eq:optimization_d},\eqref{eq:optimization_c},\eqref{eq:optimization_b} ,  \eqref{eq:location_est},  \eqref{eq:channel_est}.
\end{align}
\end{subequations}
Due to the complex nature of the functions $\mathcal{F}(\cdot)$, finding a globally optimal solution analytically for this problem is intractable. This paper proposes a data-driven approach to make this problem tractable. 
%\subsection{Architecture of Image Detection Network}
\begin{figure}[tbp]
\centerline{\includegraphics[width=1 \linewidth]{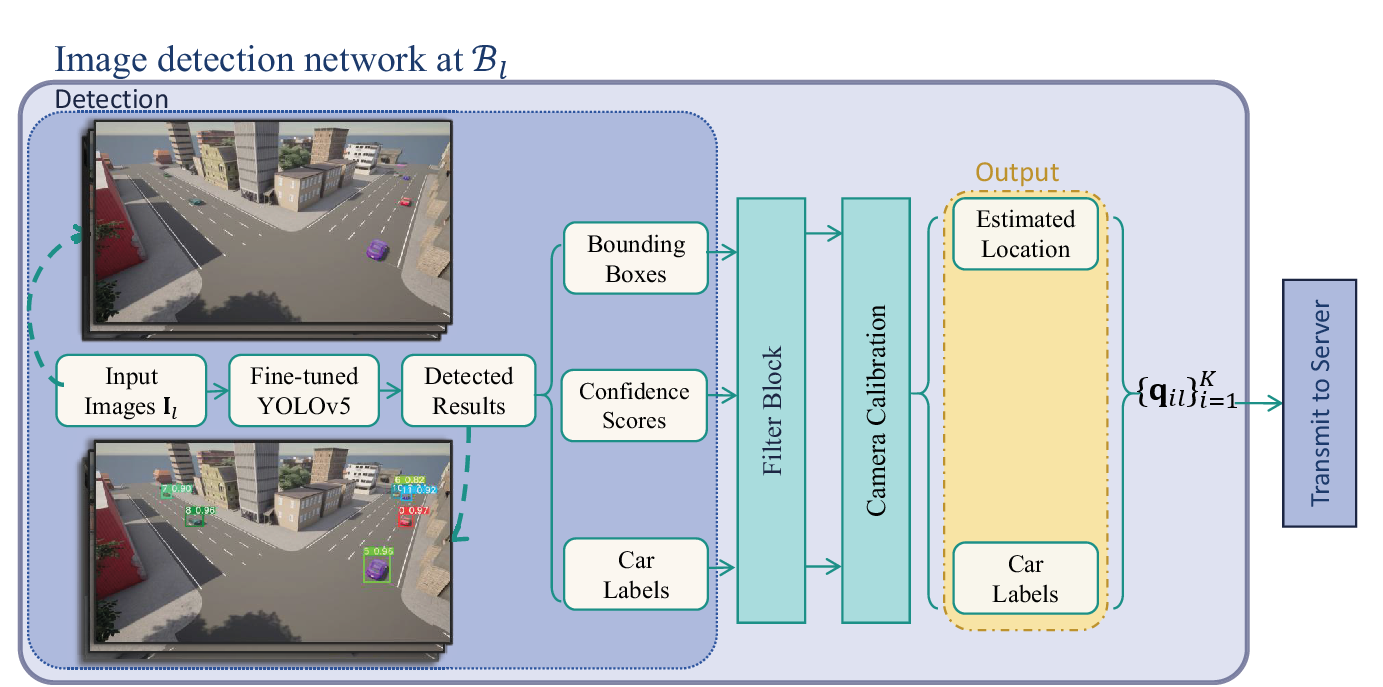}}
\caption{ Architecture of the image detection network.}
\label{Detection_NN}
\end{figure}
\section{Image-Based Location Estimation}
\label{section:imgae}
In this section, we provide details of the image detection neural network, which is utilized for mapping visual data to the estimated user locations, i.e., implementing the function $\mathcal{G}_l(\cdot)$ as defined in \eqref{eq:location_est}.
We assume that each AP is equipped with a camera to capture images of the surrounding environment. To reduce data transmission load and protect user privacy, only the estimated user locations, rather than the full image data, are transmitted to the central server for subsequent system optimization.

{In a system with $K$ users, $L$ APs, and each AP equipped with $M$ antennas, the amount of data in the transmitted pilots from the BS to the central server is $\mathcal{O}(MKL)$ complex numbers. The transmission of image-based location information introduces an additional data overhead of $\mathcal{O}(3KL)$. Typically, $M$ is much larger than 3, therefore transmitting location estimates is an efficient method.}
 
To enable local location estimation, we assume that each AP deploys an image detection neural network to estimate user locations from the captured images. As shown in Fig.~\ref{Detection_NN}, the architecture of the proposed image detection network consists of three main components:
\subsubsection{Detection Block}
The first component is the detection block, where we utilize YOLOv5 \cite{yolov5}, short for ``You Only Look Once,'' to detect the target UEs within the images.
We assume that the UEs are cars having different colors to differentiate them in the image, and each car is assigned a distinct class label.
This assumption enables us to generate a labeled dataset containing bounding boxes and class labels for each UE.  By training YOLOv5 on this dataset, the model learns to accurately detect and localize the target UEs in the images. Once trained, the parameters of YOLOv5 are fixed, and the model can process the new images to provide the detection results.
The detection results are divided into three components: bounding boxes, class labels, and confidence scores. The bounding box contains information about the target UE's center coordinates, width, and height to specify the location and size of the detected object within the image. The class labels identify different detected cars, and the confidence scores represent the probability that the detection is accurate.
\subsubsection{Filtering Block}
The second component is the filtering block. It refines detected results by filtering out those with confidence scores below 0.7, and then resolves redundancy by retaining only the detected object with the highest confidence score when multiple detected objects share the same class label. This process ensures that only the most accurate detection is used for further processing. 
\subsubsection{Camera Calibration Block} 
\begin{figure}[tbp]
\centerline{\includegraphics[width=0.8 \linewidth]{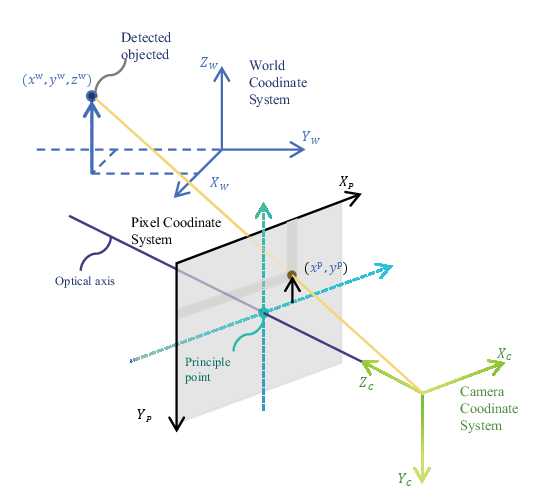}}
\caption{Three coordinate systems used in the camera calibration.}
\vspace{-5pt}
\label{calibration}
\end{figure}

The third component is the camera calibration block \cite{cali92}. Camera calibration is a commonly used method in computer vision that maps pixel points in the image to their corresponding real-world locations.  As shown in Fig.~\ref{calibration}, three coordinate systems are involved in this calibration process:  i) the world coordinate system, which defines the positions of objects in the real world; ii) the camera coordinate system, centered at the camera's optical center and influenced by the camera's position and rotation; and iii) the pixel coordinate system, where coordinates correspond to the pixel grid on the camera sensor, with its origin at the principal point on the image plane.

Let $(x^\text{w}_k, y^\text{w}_k, z^\text{w}_k)$ represent the coordinates of the center of $\mathcal{U}_k$ in the world coordinate system, and let $(x^\text{c}_{kl}, y^\text{c}_{kl}, z^\text{c}_{kl})$ denote the corresponding point in the camera coordinate system of the camera at $\mathcal{B}_l$. The transformation between the world coordinates and the camera coordinates is given by:

\begin{equation}
\label{eq:extrinsic}
\left[\begin{array}{l}
x^\text{c}_{kl}\\
y^\text{c}_{kl} \\
z^\text{c}_{kl}
\end{array}\right]=\left[\mathbf{Q}_l \mid \mathbf{T}_l\right]\left[\begin{array}{c}
x^\text{w}_k \\
y^\text{w}_k \\
z^\text{w}_k \\
1
\end{array}\right],
\end{equation}
where $\left[\mathbf{Q}_l \mid \mathbf{T}_l\right]$ is known as the extrinsic matrix. Here, $\mathbf{Q}_l \in \mathbb{R}^{3 \times 3}$ is the rotation matrix that describes the orientation of the camera relative to the world coordinate system, and $\mathbf{T}_l \in \mathbb{R}^3$ is the translation vector that specifies the position of the camera's optical center in the world coordinate system. The notation $[\cdot \mid \cdot]$ represents the column-wise concatenation of the rotation matrix and the translation vector. 

At $\mathcal{B}_l$, the functional relationship between the coordinates $(x^\text{c}_{kl}, y^\text{c}_{kl}, z^\text{c}_{kl})$ in the camera coordinate system and the corresponding pixel coordinates $(x^\text{p}_{kl}, y^\text{p}_{kl})$ in the pixel coordinate system can be represented by:
\begin{equation}
\label{eq:intrinsic}
\hbar_{kl}\left[\begin{array}{l}
x^\text{p}_{kl} \\
y^\text{p}_{kl} \\
1
\end{array}\right]=\mathbf{D}_l\left[\begin{array}{l}
x^\text{c}_{kl} \\
y^\text{c}_{kl} \\
z^\text{c}_{kl}
\end{array}\right]=\left[\begin{array}{llc}
f_l & 0 & c_{1,l} \\
0 & f_l & c_{2,l} \\
0 & 0 & 1
\end{array}\right]\left[\begin{array}{l}
x^\text{c}_{kl} \\
y^\text{c}_{kl} \\
z^\text{c}_{kl}
\end{array}\right],
\end{equation}
where $\hbar_{kl}$ is a scaling factor that aligns the pixel and camera coordinate systems and $\mathbf{D}_l$ represents the intrinsic camera matrix for camera $\mathcal{B}_l$. In the matrix $\mathbf{D}_l$, $f_l$ represents the focal length of the camera at $\mathcal{B}_l$, while $(c_{1,l}, c_{2,l})$ represent the horizontal and vertical coordinates of the principal point of the camera at $\mathcal{B}_l$.

In practical scenarios, it is reasonable to assume that the height of the cars can be estimated.  Let $\hat{z}^\text{w}_k$ denote the estimated height of the center of $\mathcal{U}_k$ in the world coordinate system.
Based on (\ref{eq:extrinsic}) and (\ref{eq:intrinsic}), the estimated location of the $k$-th car, $[\hat{x}^\text{w}_{k}, \hat{y}^\text{w}_{k}]$, can be derived from the pixel coordinates $(x^\text{p}_{kl}, y^\text{p}_{kl})$ by substituting the real height $z^\text{w}_k$ with the $\hat{z}^\text{w}_k$. This estimation follows from (\ref{eq:estimation}) at the bottom of the page, where $\mathbf{F}_l = \mathbf{D}_l \times \left[\mathbf{Q}_l \mid \mathbf{T}_l\right]$, and $[\mathbf{F}_l]_{m,n}$ indicates the element in the $m$-th row and $n$-th column of the matrix $\mathbf{F}_l$.

After the images are processed by the detection, filtering, and calibration blocks within the image detection network, the AP transmits the obtained location information to the server.
\begin{strip}
\vspace{2pt}
	\hrulefill
	\begin{equation}
		\left[\begin{array}{l}
			\hat{x}^\text{w}_{kl} \\
			\hat{y}^\text{w}_{kl}
		\end{array}\right] =
		\left[\begin{array}{cc}
			x^\text{p}_{kl} [\mathbf{F}_l]_{3,1} - [\mathbf{F}_l]_{1,1}, & x^\text{p}_{kl} [\mathbf{F}_l]_{3,2} - [\mathbf{F}_l]_{1,2} \\
			y^\text{p}_{kl} [\mathbf{F}_l]_{3,1} - [\mathbf{F}_l]_{2,1}, & y^\text{p}_{kl} [\mathbf{F}_l]_{3,2} - [\mathbf{F}_l]_{2,2}
		\end{array}\right]^{-1} \cdot
		\left[\begin{array}{c} 
			\hat{z}^\text{w}_k[\mathbf{F}_l]_{1,3} + [\mathbf{F}_l]_{1,4}-\hat{z}^\text{w}_k x^\text{p}_{kl} [\mathbf{F}_l]_{3,3} - x^\text{p}_{kl} [\mathbf{F}_l]_{3,4} \\
			\hat{z}^\text{w}_k[\mathbf{F}_l]_{2,3} + [\mathbf{F}_l]_{2,4}-\hat{z}^\text{w}_k y^\text{p}_{kl} [\mathbf{F}_l]_{3,3} - y^\text{p}_{kl} [\mathbf{F}_l]_{3,4} 
		\end{array}\right].
		\label{eq:estimation}
	\end{equation}
\end{strip}

\section{Proposed GNN Framework for Solving User Association and Beamforming}
\label{section:GNN}

The user association and beamforming optimization problem \eqref{eq:optimization}
follows certain permutation equivariance property, i.e., the permutation of the users' 
locations or users' CSI would result in the same permutation of the optimal $a_{kl}$ 
and $\mathbf{v}_{kl}$. This motivates the use of GNN as a suitable neural network
architecture for user association and beamforming optimization. 
In a wireless network, the user association and beamforming for a user can
inherently affect and be affected by the corresponding decisions made for other users.
This paper proposes a GNN-based archiecture to map the obtained information (the estimated locations from images or the received pilots) to the designed beamforming and user association parameters. GNNs are adopted in this paper, because they are well-suited for optimization problems with the permutation equivariance property and excel at capturing complex relational dependencies between nodes (UEs) in the graph.

We introduce three GNNs to solve the user association and beamforming problem:
\begin{itemize}
	\item GNN-I, which uses image based location estimation as input 
		to produce a set of candidate user association and beamforming patterns; 
	\item GNN-P, which uses received pilots as input to 
		 produce another set of candidate user association and beamforming patterns; 
	\item GNN-C, which utilizes features from GNN-I and features from GNN-P as input to produce the final optimized set of user association and beamforming patterns. 
\end{itemize}
%{ \subsubsection{GNN based on image-derived location estimates (GNN-I)} GNN-I utilizes estimated locations obtained from images as the input to the GNN to optimize system configurations.
%{ \subsubsection{GNN based on received pilots (GNN-P)} GNN-P utilizes processed received pilots as the input to the GNN to optimize system configurations.
%{ \subsubsection{GNN based on combined image and pilot data (GNN-C)} GNN-C utilizes multimodal data (node features from GNN-I and node features from GNN-P) as input to optimize system configurations.
  \begin{figure*}[tbp]
\centerline{\includegraphics[width=0.87 \linewidth]{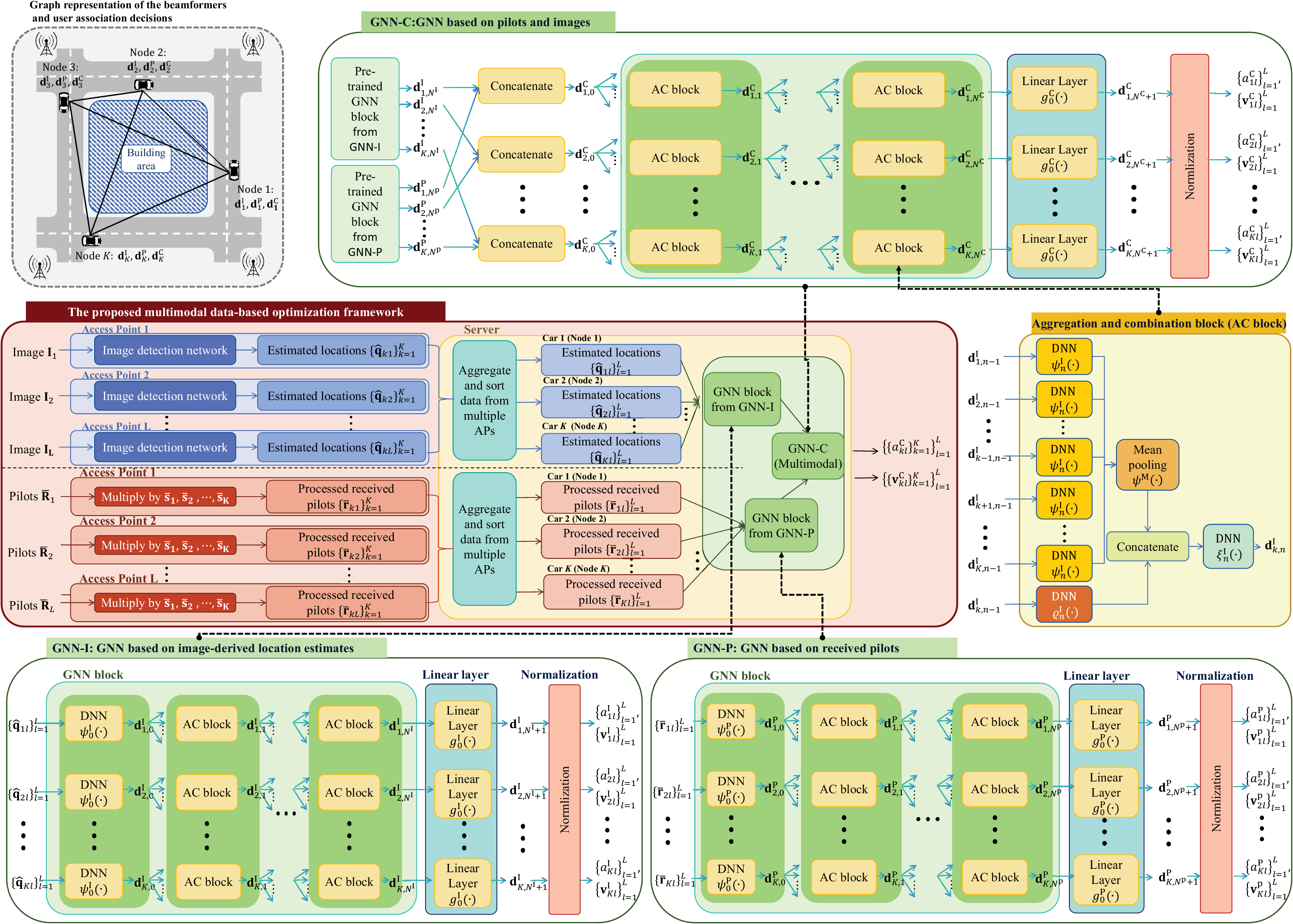}}
\caption{ Architecture of the proposed GNN network.}
\label{structure}
\end{figure*}

{GNN-I and GNN-P are trained separately using their respective training data and can be directly utilized when only one type of data is available. For the training of GNN-C with multimodal data, we first ensure that GNN-I and GNN-P are well-trained independently to guarantee robust feature extraction from two different data sources. Then, GNN-C leverages the multimodal features extracted from the well-trained GNN-I and GNN-P to optimize user association and beamforming.}

As illustrated in Fig.~\ref{structure}, we model the relationships among users using a graph, where $K$ nodes represent the $K$ users in the network, and the edges indicate that the connected nodes are interdependent and can exchange information during the feature update process. 
In the optimization problem, a fully connected graph is utilized to model the relationships among the users, because the configurations of the user association and beamforming are interdependent across the users and the APs. The $k$-th node in the graph is associated with the updated feature vector of $\mathcal{U}_k$. We utilize $\mathbf{d}^{\text{I}}_{k,n}$, $\mathbf{d}^{\text{P}}_{k,n}$, and $\mathbf{d}^{\text{C}}_{k,n}$ to represent $\mathcal{U}_k$'s feature vector after the $n$-th update from GNN-I, GNN-P, and GNN-C, respectively.

\subsection{GNN Based on Location Information Derived from Images}
The main goal of GNN-I is to refine the features $\mathbf{d}^{\text{I}}_{i,n}, \forall n$, to capture the necessary information aligned with the objective of the optimization.
In this section, we first introduce the input features for GNN-I. Next, we describe the three main components of GNN-I, which are designed for mapping the input features to the optimized configurations. Finally, we discuss the training and inference processes.
\subsubsection{Input Features}
As shown in Fig.~\ref{structure}, each AP is equipped with an image detection neural network that processes the captured images to estimate the locations of UEs. The estimated locations are transmitted to a central server, where they are aggregated and sorted according to the class labels of the UEs. For the GNN, the input features of the $k$-th node correspond to the estimated locations of $\mathcal{U}_k$ across all APs, represented as $\{\hat{\mathbf{q}}_{kl}\}_{l=1}^L$.
\subsubsection{GNN Block}
The GNN block takes features extracted from images as input and refines node features by aggregating information from neighboring nodes to realize more effective user association and beamforming optimization.
The input $\{\hat{\mathbf{q}}_{kl}\}_{l=1}^L$ of the $k$-th node is provided to a function $\psi^\text{I}_0(\cdot)$ to produce the initial node features $\mathbf{d}_{k,0}$, i.e.,
\begin{equation}
\mathbf{d}^\text{I}_{k,0}=\psi^\text{I}_0\left(\{\hat{\mathbf{q}}_{kl}\}_{l=1}^L\right),
\end{equation}
where $\psi^\text{I}_0(\cdot)$ denotes a fully connected network (FCN) with $N^\text{I}_0$ layers and each layer is followed by a batch normalization (BN) layer and a rectified linear unit (ReLU) activation function.
The initial features $\mathbf{d}^\text{I}_{k,0}$ are then processed through a series of aggregation and combination (AC) blocks, where each stage updates the node features by integrating information from neighboring nodes. Let $\mathbf{d}^{\text{I}}_{k,n}$ denote the output features of the $k$-th node after the $n$-th AC block. The features are updated according to the following equation:
\begin{equation}
\mathbf{d}^\text{I}_{k,n}=\xi_n\left(\psi^\text{M} \left( \left\{\psi^\text{I}_n(\mathbf{d}^\text{I}_{i,n-1})
\right\}_{i=1,i \neq k}^{K} \right), \varrho_n^\text{I}\left(  \mathbf{d}^\text{I}_{k,n-1}\right)  \right),
\end{equation}
where  $\psi^\text{M}(\cdot)$ is an element-wise mean pooling function. The functions $\psi^\text{I}_n(\cdot)$, $\varrho_n^\text{I}(\cdot)$, and $\xi^\text{I}_n(\cdot)$ represent fully connected networks (FCNs) with $N^\text{I}_{n,1}$, $N^\text{I}_{n,2}$, and $N^\text{I}_{n,3}$ layers, respectively. Each fully connected layer is followed by a BN layer and a ReLU activation function.  The neural networks $\psi^\text{I}_n(\cdot)$ and $\varrho_n^\text{I}(\cdot)$ are applied to extract information from the features of neighboring nodes and the node itself, respectively. The extracted information is concatenated and subsequently processed by $\xi_n(\cdot)$, which integrates the combined data to produce updated node features $\mathbf{d}^\text{I}_{k,n}$.
To ensure sufficient information change and extraction, the AC block refines the initial features for $N^\text{I}$ times, resulting in the updated node features $\mathbf{d}^\text{I}_{k,1}, \mathbf{d}^\text{I}_{k,2}, \ldots, \mathbf{d}^\text{I}_{k,N^\text{I}}$.
\subsubsection{Linear Layer Block}
After $N^\text{I}$ updates, the outputs of the $k$-th node $\mathbf{d}^\text{I}_{k,N^\text{I}}$ are fed into a linear function $\mathbf{g}^\text{I}_0(\cdot)$. The linear layer is designed to transform the node features to a fixed length vector $\mathbf{d}^\text{I}_{k,N^\text{I}+1}\in\mathbb{C}^{ML+L} $. The first $L$ elements of $\mathbf{d}^\text{I}_{k,N^\text{I}+1}$ represent the user association decisions for $\mathcal{U}_k$ and are denoted as $ \{\tilde{a}^\text{I}_{kl}\}_{l=1}^L $. The last $ML$ elements of  $\mathbf{d}^\text{I}_{k,N^\text{I}+1}$ is defined as $\{{\tilde{\mathbf{v}}^\text{I}_{kl}}\}^L_{l=1}$, corresponding to the beamforming designed for $\mathcal{U}_k$.
\subsubsection{Normalization Block}
This block is designed to map the output $\tilde{a}^\text{I}_{kl}$ and $\tilde{\mathbf{v}}^\text{I}_{kl}$ from the linear layer to the normalized outputs that satisfy the constraints \eqref{eq:optimization_d}, \eqref{eq:optimization_c} and \eqref{eq:optimization_b}.
%In the deep learning framework, discrete constraint such as \eqref{eq:optimization_d} can hinder gradient backpropagation, negatively impacting the training process\cite{liu2024bridging}. Therefore, the constraint \eqref{eq:optimization_d} is not considered here and will be addressed in the following section. 
%In the normalization block, we modify the original constraint \eqref{eq:optimization_d}, which required $a_{kl}$ to be either 0 or 1, to a relaxed form:
%\begin{equation}
%\label{eq:constraint_relax}
%0 \leq a_{kl} \leq 1,
%\end{equation}
First, the user association $\hat{a}^\text{I}_{kl}$ and beamforming vector $\hat{\mathbf{v}}^\text{I}_{kl}$ can be normalized via the equation:
\begin{equation}
\hat{a}^\text{I}_{kl}=\frac{|\tilde{a}^\text{I}_{kl}| }{\sum_{l=1}^L |\tilde{a}^\text{I}_{kl}|},  \ 	\hat{\mathbf{v}}^\text{I}_{kl}=\sqrt{\bar{p}_l}\frac{\tilde{\mathbf{v}}^\text{I}_{kl} } {\sqrt{\sum_{i=1}^K \|\tilde{\mathbf{v}}^\text{I}_{il}  \|^2_2}},
\end{equation}
where the obtained $\hat{a}^\text{I}_{kl}$ and $\hat{\mathbf{v}}^\text{I}_{kl}$ satisfy the  \eqref{eq:optimization_c} and \eqref{eq:optimization_b}.
Finally, we quantize the maximum of the elements in $\{\hat{a}^\text{I}_{kl}\}_{l=1}^L$ as one and others as zeros, in order to ensure that it satisfies the binary constraints \eqref{eq:optimization_d}, as follows:
% . The ideal quantization of $\hat{a}_{kl}$, denoted by $\hat{a}^\star_{kl}$, can be expressed as
\begin{equation}
\hat{a}^{\text{I}\star}_{kl} =
\begin{cases}
1, & \text{if} \ l = \underset{j}{\operatorname{argmax}} \left( \hat{a}^\text{I}_{kj} \right), \\
0, & \text{otherwise.}
\end{cases}
\end{equation}
The quantization process ensures that each user is associated with exactly one AP. However, quantization is a non-differentiable function, which poses challenges for backpropagation-based neural network learning.
The STE modification can be used to solve this problem. The STE enables gradient-based learning by allowing non-differentiable discrete variables, such as user associations, to be treated as differentiable during the backpropagation. This can be realized by
\begin{equation}
 {a}^\text{I}_{kl}  = \hat{a}^\text{I}_{kl}  + \text{stop\_grad}(\hat{a}^{\text{I}\star}_{kl} - \hat{a}^\text{I}_{kl} ) ,
\end{equation}
where $\text{stop\_grad}(\cdot)$ stops the gradient from being computed during backpropagation for the enclosed expression but still passes through the forward values during forward propagation. 

Observe from \eqref{eq:rate} that the user association decisions $a^\text{I}_{kl}$ can be thought of as a binary coefficient of the beamforming vectors $\mathbf{v}^\text{I}_{kl}$. Thus, if $a^\text{I}_{kl}$ is set as 0 at any point in the learning process, $\mathbf{v}^\text{I}_{kl}$ would be excluded from subsequent gradient calculation, thus preventing updates to the beamforming vector for that specific user-AP pair.  This can impede the learning process, because no further updates can occur for such user-AP pairs.
%, which may affect the model's overall performance in the training. 
To address this issue, we propose to update the user association decision $a^\text{I}_{kl}$ as follows:
\begin{align}
 {a}^\text{I}_{kl}  &= (1-\lambda) \cdot \hat{a}^\text{I}_{kl}  +  \lambda  \cdot \left(\hat{a}^\text{I}_{kl}+ \text{stop\_grad}(  \hat{a}^{\text{I}\star}_{kl} - \hat{a}^\text{I}_{kl} ) \right) \nonumber \\
 &= \hat{a}^\text{I}_{kl} + \lambda \cdot \text{stop\_grad}(  \hat{a}^{\text{I}\star}_{kl} - \hat{a}^\text{I}_{kl} ),
\end{align}
where $\lambda$ is a balancing coefficient that controls the trade-off between continuous learning and the binary output. By introducing $\lambda$, we allow for a gradual transition between continuous learning and binary decisions, ensuring that the model adjusts and updates beamforming vectors $\mathbf{v}^\text{I}_{kl}$ during early epochs of training. As $\lambda$ increases over time, the system moves closer to binary outputs, progressively locking in final user associations. This gradual transition mechanism prevents the premature exclusion of user-AP pairs from the optimization process and ensures that the beamforming vectors are jointly optimized with the user association decisions.

\subsubsection{Neural Network Training}
Since \eqref{eq:optimization} is a nonconvex problem with discrete constraints, obtaining labeled data for training GNN can be challenging and costly. %We employ unsupervised learning to enable the model to optimize its parameters without labeled data.
Given the normalized $\hat{\mathbf{v}}^\text{I}_{kl}$ and $ {a}^\text{I}_{kl}$ from GNN, we can obtain the estimated rate $\hat{R}^\text{I}_k $ from (\ref{eq:rate}).
By defining the negative of the objective function as the loss, the GNN model can learn directly from the input data by optimizing the objective function given in (\ref{eq:optimization_a}).
Additionally, to prevent the learned user association from being 0 in the initial epochs, we introduce a penalty term $\sum_{kl}({a}^\text{I}_{kl}-\frac{1}{L})^2$ that prevents values from being directly set to 1.
The loss function can be formulated as $-\mathbb{E}\left[\min_{1 \leq k \leq K} \hat{R}^\text{I}_k  \right]+\lambda_p \sum_{kl}({a}^\text{I}_{kl}-\frac{1}{L})^2  $, where the expectation $\mathbb{E}[\cdot]$ represents the average over the training batch, and $\lambda_p$ is a penalty coefficient that decreases over time to allow for smoother transitions during optimization.

\subsubsection{Neural Network Inference}
During the inference phase, as gradient backpropagation is not required, we utilize $\hat{a}^{\text{I}\star}_{kl}$ as the final user association decision to make the output satisfy the binary constraint.
\subsection{GNN Based on Received Pilots}
As shown in the Fig.~\ref{structure}, the GNN-P leverages RF pilots to optimize beamforming $\mathbf{v}^\text{P}_{kl}$ and user association $a^\text{P}_{kl}$. The training and inference process of GNN-P are exactly the same as GNN-I. The structure of GNN-P is shown in Fig.~\ref{structure}, where $\mathbf{d}^\text{P}_{k,n}$ represents the node features at the $n$-th layer for $\mathcal{U}_k$. %From \cite{tao}, we know a deep neural network is capable of parameterizing the mapping from the received pilots to an optimized system configuration without channel estimation. Bypassing the channel estimation, we directly use the processed received pilots  $\bar{\mathbf{r}}_{kl}$ based on equation (\ref{eq:pilot2}). 

%During the interference phase, since gradient backpropagation is not required, we just utilize the $\hat{a}_{il}^{\star}$ as the final association decisions to make the output satisfy the binary constraint.
%In GNN-P, the loss function is defined as $ -\mathbb{E}\left[\min _{1 \leq k \leq K} \hat{R}_k\right] + \lambda_p \mathbb{E}\left[\sum_{il}\left(\hat{a}_{il} - \hat{a}_{il}^{\star}\right)^2\right] $, where the second term is used to encourage binary association decisions, and $\lambda_p$ is an hyperparameter that balances the trade-off between the two terms in the loss function.

\subsection{GNN Based on Multimodal Data}
 
As shown also in Fig.~\ref{structure}, the GNN-C leverages both image and pilot data to optimize beamforming $\mathbf{v}^\text{C}_{kl}$ and user associaiton $a^\text{C}_{kl}$.
When GNN-I and GNN-P are well-trained, their node features capture essential features from the images and received pilots, respectively, based on the objective of the system optimization. 
We propose to use the concatenation of the node features from GNN-I and GNN-P as input to GNN-C. The structure, training, and inference process of GNN-C are exactly the same as GNN-I and GNN-P. The structure of GNN-C is shown in Fig.~\ref{structure}, where $\mathbf{d}^\text{C}_{k,n}$ represents the node features at the $n$-th layer for $\mathcal{U}_k$.

{In this paper, we only consider the permutation equivariance property of the users, because UE distribution is typically dynamic and more complex than that of APs. In the user association and beamforming problem, the APs also exhibit permutation equivariance properties. When the number of APs is large and the network becomes more complex, the GNN structure proposed in \cite{Yunqi24,Jia22} can be utilized for the APs for better adaptation.}

\section{Simulation Results}
In this section, we generate the dataset and evaluate the performance of the proposed deep learning method.
\begin{figure*}[htb]
 \centerline{\includegraphics[width=0.95 \linewidth]{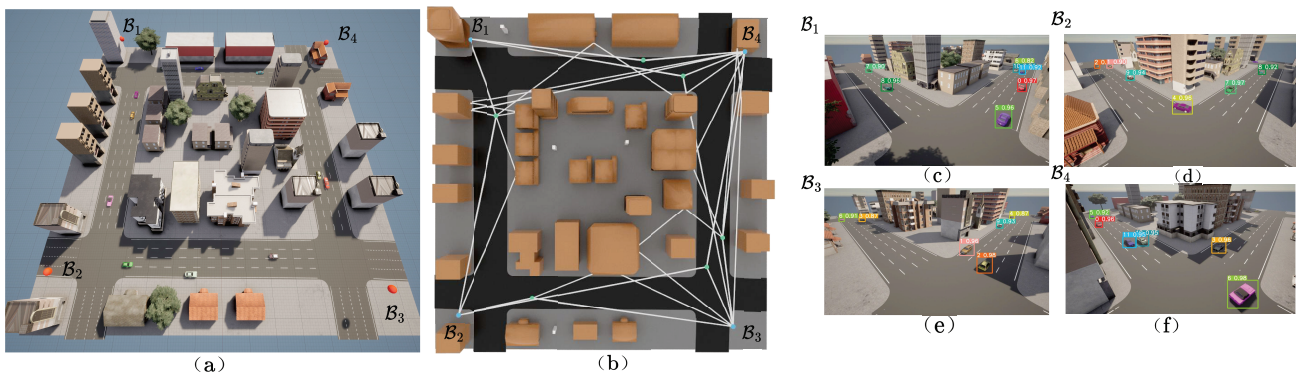}}
\caption{(a) The 3D model of the simulation environment in the CARLA simulator. 
    (b) The 3D model of the environment recreated in the Sionna simulator. 
    (c)-(f) Visualization of the detection results for $\mathcal{B}_1$ to $\mathcal{B}_4$, respectively.}
\label{simulation_scenario}
\end{figure*}
\subsection{Image Generation}
% \begin{figure}[tbp]
%\centerline{\includegraphics[width=1 \linewidth]{figs/carla_scenario.pdf}}
%\caption{The 3D model of the scenario.}
%\label{simulation_scenario}
%\end{figure}
We utilize the open-source simulator for autonomous driving, CARLA \cite{carla}, to construct the simulation environment. As shown in Fig.~\ref{simulation_scenario}(a), the scenario covers an area of 150 meters by 150 meters. The running area for the cars is confined to a smaller square road with a length of approximately 110 meters. In the simulation, four APs are positioned at the four corners of the intersection, which is represented by the four orange spheres in the figure. These APs are positioned at the coordinates $\mathcal{B}_1$ [60, -55, 15], $\mathcal{B}_2$ [-55, -60, 15], $\mathcal{B}_3$ [-60, 55, 15], and $\mathcal{B}_4$ [55, 60, 15].
Each AP is equipped with a camera, and all APs are strategically positioned to ensure coverage of two intersecting roads. The orientations for the cameras are set as follows: $[-30^\circ, -224^\circ, 0^\circ]$ for $\mathcal{B}_1$, $[-30^\circ, -315^\circ, 0^\circ]$ for $\mathcal{B}_2$, $[-30^\circ, -43^\circ, 0^\circ]$ for $\mathcal{B}_3$, and $[-30^\circ, -135^\circ, 0^\circ]$ for $\mathcal{B}_4$, where three elements represent the pitch, roll, and yaw angles, respectively. All cameras have a 100-degree field of view (FOV) and capture images at a resolution of 1280x720 pixels. From Fig.~\ref{simulation_scenario}(c-f), we observe that each camera provides a clear view of two intersecting roads, ensuring comprehensive coverage of the area of interest.

The simulation utilizes cars to represent the UEs, with each car assigned a unique color for identification. The cars are randomly generated at various spawn points on the roads and are assigned random speeds to ensure a realistic variation in traffic flow. Four cameras at 4 APs are positioned to take pictures simultaneously at randomly selected time intervals during the simulation. 
\subsection{Channel Generation}
In this simulation, we utilize Sionna \cite{sionna}, an open-source ray tracing software, to model and to analyze the propagation of wireless signals in the simulation environment. The system operates at a carrier frequency of 3.5 GHz. The transmit power of the APs is set to 25 dBm, and the transmit power of the vehicles is set to 5 dBm. The noise power is set at -55 dBm for both the APs and the UEs. The APs are equipped with planar array antennas with $M_r=4$ rows and $M_c=4$ columns.

We import simplified representations of buildings, roads, ground, and trees into Sionna to create a realistic yet simplified simulation environment, as shown in Fig.~\ref{simulation_scenario}(b). The influence of leaves of trees is omitted to simplify the model. The blue points represent the APs, and the green points represent the cars. Sionna allows for accurate simulation of complex environments by taking into account various materials and their electromagnetic properties. The materials used in our simulation are shown in Table \ref{tab:scenario-settings}. The antennas of the UEs are assumed to be placed at a fixed height of 0.4 meters.
\begin{table}[t]
\vspace{-2pt}
    \caption{Scenario Settings for Sionna Simulation}
    \centering
    \begin{tabular}{|l|l|}
        \hline
        \textbf{Component} & \textbf{Setting} \\
        \hline
        Building Material & itu\_concrete \\
        Road Material & itu\_very\_dry\_ground \\
        Ground Material & itu\_medium\_dry\_ground \\
        Trunk Material & itu\_wood  \\
        \hline
    \end{tabular}
    \vspace{-2pt}
    \label{tab:scenario-settings}
\end{table}

We obtain the complex channel coefficients, angles of arrival, and angles of departure for each multipath component from Sionna.
Let $N^R_{kl}$ represent the number of paths from $\mathcal{U}_k$ to $\mathcal{B}_l$. Let $A^R_{kl}(n)$ denote the complex channel coefficient to be multiplied with the $n$-th multipath component. Let $\phi^R_{kl}(n)$ and $\theta^R_{kl}(n)$ denote the azimuth and zenith departure angle of the $n$-th multipath component from $\mathcal{B}_l$, respectively.
For a planar array, the elements experience different phase shifts based on their spatial positions. We define functions $v_1(m, M_c)$ and $v_2(m, M_c)$ to map the element index $m$ to the respective row and column positions in the planar array, centered around the array's midpoint:
\begin{align}
  v_1(m, M_c) &= \left\lfloor \tfrac{m}{M_c} \right\rfloor - \tfrac{M_r - 1}{2}, \\
  v_2(m, M_c) &=\operatorname{mod}(m, M_c) - \tfrac{M_c - 1}{2}.
\end{align}
Define the intermediate variables for clarity:
\begin{align}
  \alpha_{kl}^R(n, m) &= v_1(m, M_c) \sin(\theta^R_{kl}(n)) \cos(\phi^R_{kl}(n)), \\
  \beta_{kl}^R(n, m) &= v_2(m, M_c) \cos(\theta^R_{kl}(n)),
\end{align}
where $\alpha_{kl}^R(n, m)$ and $\beta_{kl}^R(n, m)$ denote the phase shifts introduced by the row and column positions, respectively. The complex channel coefficient for the $m$-th array element is given as a summation over all multipath components:
\begin{align}
  [\mathbf{h}_{kl}]_m = \sum\limits_{n} A^R_{kl}(n) e^{ \frac{2 j \pi d_{\text{R}}}{\lambda_c} \left( \alpha_{kl}^R(n, m) + \beta_{kl}^R(n, m) \right) },
\end{align}
where $\lambda_c$ denote the wavelength, and $d_{\text{R}}$ represent the spacing between the antenna elements. In this setup, $d_{\text{R}}$ is set as half of the wavelength $\lambda_c/2$.
The channel coefficients are utilized to train the proposed GNN under realistic conditions.
\subsection{Neural Network Configurations}
This section introduces the proposed neural network configurations and training setups.

For the image detection framework, we employ YOLOv5s for object detection tasks. The CARLA simulator is used to generate a dataset with bounding boxes and labels, consisting of 16,000 images for training and 200 images for validation. The training epoch is set as 200. After training, the YOLOv5s parameters are fixed and applied to subsequent tasks.

For the GNN framework, the number of AC blocks in GNN-I, GNN-P, and GNN-C  is set to 2,  i.e., $N^\text{I}=N^\text{P}=N^\text{C}=2$. The architecture and parameters of the FCNs for GNN-I, GNN-P and GNN-C are summarized in Table \ref{tab:gnn_parameters}. 
The Adam optimizer is employed for all GNNs, with a learning rate set to 0.0002. The batch sizes for GNN-I, GNN-P, and GNN-C are 512, 1024, and 1024, respectively. 
The dataset generated using CARLA and Sionna is shared across three GNN models, with a training set comprising $1.97 \times 10^5$ samples, a validation set of $1.8 \times 10^3$ samples, and a test set consisting of 200 samples.
%In the implementation, $\lambda$ is dynamically adjusted to enable a smooth transition toward binary outputs, following the formula: determined by:
%\begin{equation}
% \lambda= \min(\text{epoch}/100,1).
%\end{equation}
%By epoch 100, $\lambda$ reaches 1, solidifying the user association decisions as binary.
\begin{table*}[ht]
\renewcommand{\arraystretch}{1.4}
\centering
\caption{Architecture and Parameters of the Neural Networks (GNN-I, GNN-P, GNN-C)}
\label{tab:gnn_parameters}
\resizebox{0.8 \textwidth}{!}{ 
\begin{tabular}{|cc|cc|cc|}
	\hline
    \multicolumn{2}{|c|}{GNN-I} & \multicolumn{2}{|c|}{GNN-P} & \multicolumn{2}{|c|}{GNN-C} \\
	\hline
	Name & Size & Name &  Size& Name & Size   \\
    \hline
    $\psi^\text{I}_0$ & $(3L\times 512 , 512 \times512) $ & $\psi^\text{P}_0$ & $(2LM\times 1024 , 1024 \times 1024)$   & $\psi^\text{C}_0$ & $(1536 \times 1024, 1024 \times 1536)$   \\
      \hline    
      \makecell{      $\psi^\text{I}_1$, $\varrho^\text{I}_1 ,\psi^\text{I}_2$, $\varrho^\text{I}_2$ } &$(512 \times512 ,  512 \times 512)$       &  \makecell{$\psi^\text{P}_1$, $\varrho^\text{P}_1, \psi^\text{P}_2$, $\varrho^\text{P}_2$, } &$(1024 \times 1024 ,  1024 \times 1024)$  & \makecell{$\psi^\text{C}_1$, $\varrho^\text{C}_1,\psi^\text{C}_2$, $\varrho^\text{C}_2$, } &$(1536 \times 1024,  1024 \times 1536)$  \\
      \hline
       $\xi^\text{I}_1$, $\xi^\text{I}_2$& $(1024 \times 512 , 512\times 512)$ & $\xi^\text{P}_1$, $\xi^\text{P}_2$& $(2048 \times 1024, 1024 \times 1024)$& $\xi^\text{C}_1$, $\xi^\text{C}_2$& $(3072 \times 1024, 1024 \times 1536)$ \\
       \hline
       $g^\text{I}_0$& $512 \times (2ML+L)  $ &  $g^\text{P}_0$& $1024 \times (2ML+L)  $      &  $g^\text{C}_0$& $1536 \times (2ML+L)  $ \\
	\hline
    \multicolumn{2}{|c|}{GNN-I with varying K} & \multicolumn{2}{|c|}{GNN-P with varying K} & \multicolumn{2}{|c|}{GNN-C with varying K} \\
	\hline
	Name & Size & Name &  Size& Name & Size   \\
    \hline
    $\psi^\text{I}_0$ & $(3L\times 512,512 \times 512,512 \times 256)$ & $\psi^\text{P}_0$ & $(2LM\times 1024 ,1024\times 1024)$   & $\psi^\text{C}_0$ & $(1536 \times 2048, 2048 \times 1024)$   \\
      \hline    
      \makecell{      $\psi^\text{I}_1$, $\varrho^\text{I}_1, \psi^\text{I}_1$, $\varrho^\text{I}_2$ } &$(256 \times 512, 512 \times 512 ,512 \times 256)$       &  \makecell{$\psi^\text{P}_1$, $\varrho^\text{P}_1 , \psi^\text{P}_2$, $\varrho^\text{P}_2$, } &$(1024 \times 1024,1024 \times 1024)$  & \makecell{$\psi^\text{C}_1$, $\varrho^\text{C}_1, \psi^\text{C}_2$, $\varrho^\text{C}_2$, } &$(1024 \times 2048,2048 \times 1024)$  \\
      \hline
       $\xi^\text{I}_1$, $\xi^\text{I}_2$& $(512 \times 512,512 \times 256)$ & $\xi^\text{P}_1$, $\xi^\text{P}_2$& $(2048 \times 1024 ,1024\times 1024)$& $\xi^\text{C}_1$, $\xi^\text{C}_2$& $(2048 \times 2048,2048 \times 1024)$ \\
       \hline
       $g^\text{I}_0$& $256 \times (2ML+L) $ &  $g^\text{P}_0$& $1024 \times (2ML+L) $      &  $g^\text{C}_0$& $1536 \times (2ML+L) $ \\
          \hline
   \end{tabular}
   }
\end{table*}

  \subsection{ Benchmarks}
We compare the proposed method with the following benchmarks. The necessary statistics for the linear minimum mean square error (LMMSE) estimator are derived from 10,000 channel realizations. Additionally, all learning-based methods are designed with a carefully optimized structure to ensure a fair comparison:
 \subsubsection{Exhaustive Search with Perfect CSI} 
Given perfect CSI, we perform an exhaustive search to evaluate all possible user association configurations. For each configuration, the problem reduces to the multipoint downlink beamforming problem, which can be solved using the method proposed in \cite{multipoint}.  The configuration with the highest minimum SINR is then selected as the final result.
\subsubsection{Perfect-CSI-Based GNN}
This method employs the proposed GNN structure with perfect CSI as input, providing an upper bound on the minimum rate performance achievable by the proposed GNN structure.
%The GNN uses perfect CSI as input to fully capture the wireless environment's characteristics, providing an upper bound on the minimum rate performance for the proposed GNN structure.
{\subsubsection{Perfect-Location-Based GNN}
This method employs the proposed GNN structure with perfect location as input. It is used to validate the effectiveness of the proposed image-based location estimation neural network.}
\subsubsection{Pilot-Based GNN and Imaged-Based GNN}
 The GNN-P utilizes the RF pilots $\overline{\mathbf{r}}_{kl}$ as input. The GNN-I utilizes estimated location $\hat{\mathbf{q}}_{kl}$ from images as input. Instead of explicitly estimating the CSI and then optimizing the problem, the GNN is designed to directly optimize the proposed problem based on the input.
{\subsubsection{Direct Multimodal GNN}
This method employs a single GNN structure that directly takes all available data as input, including the estimated location and pilot data, without the separate feature extraction process~\cite{gnn_zhang}. It serves as a baseline to evaluate the advantages of using two separate GNNs for feature extraction prior to multimodal fusion.} 
\subsubsection{Exhaustive Search with LMMSE Channel Estimation}The channel is estimated using the LMMSE method. Based on the estimated channel, we perform an exhaustive search to evaluate all possible user association configurations. For each configuration, the multipoint downlink beamforming problem is solved using the approach described in \cite{multipoint}. The best configuration is selected as the final solution.
\subsubsection{Strongest-Channel-Based Assignment with LMMSE Channel Estimation}
We use the LMMSE method for channel estimation. Based on the estimated channel, each user is associated with the AP that provides the highest channel gain. After determining the user association, the problem is solved using the approach from \cite{multipoint}.
\subsection{Performance Evaluation of Image Detection Network}
 \begin{figure}[tbp]
\centerline{\includegraphics[width=0.95 \linewidth]{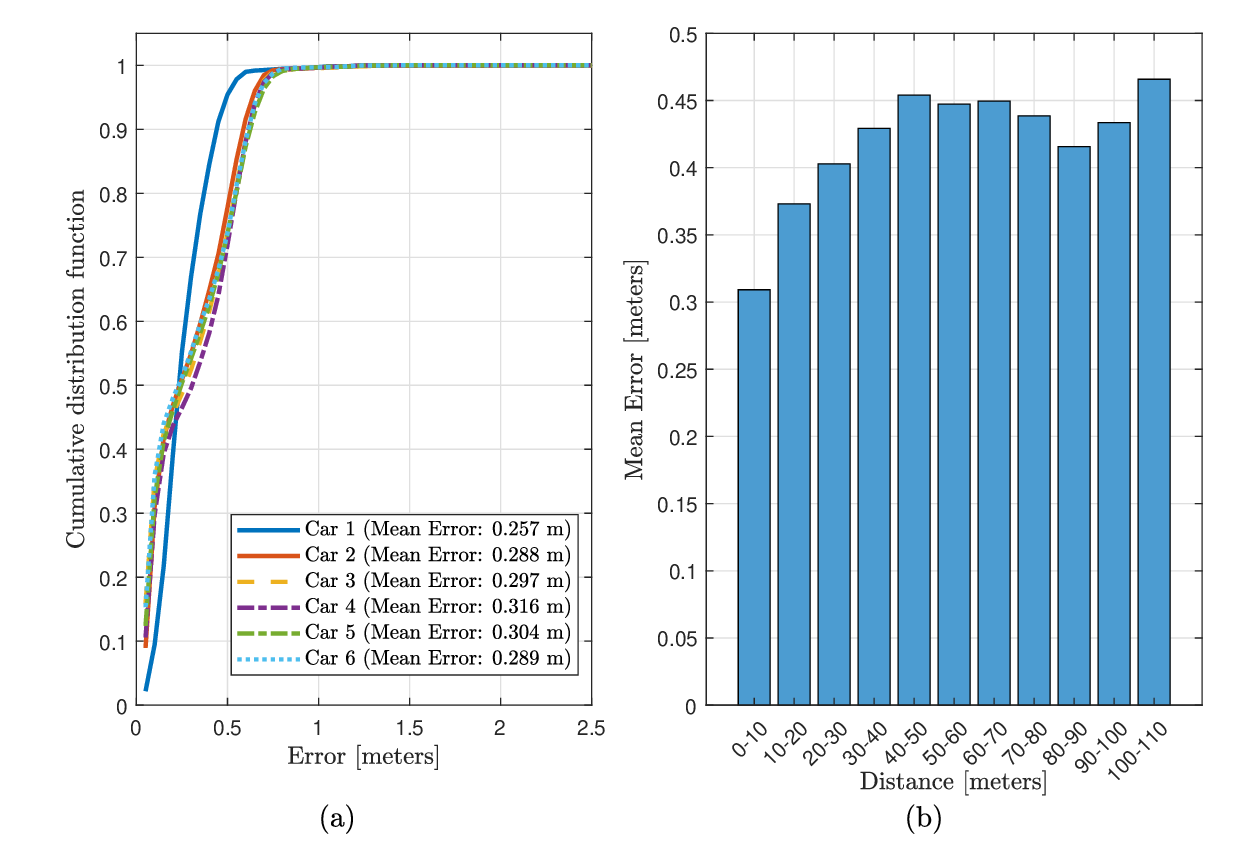}}
\caption{Errors of estimated locations from the image detection network: (a) cumulative distribution function (CDF) for estimated location errors of cars, (b) mean localization error vs. different distance ranges.}
\label{cdf}
\end{figure}
To demonstrate the performance of the detection based neural network, Fig.~\ref{cdf}(a) presents the cumulative distribution function (CDF) of location estimation errors for different cars. The data is averaged over 10,000 images.
The CDF curves demonstrate that 70\% of the location errors are smaller than 0.5 meters, and more than 99\% of the errors are smaller than 1 meter. The location error CDF of Car 1 differs from the other cars due to its bright color, which improves detection accuracy. The CDF of other cars exhibits similar location error performance.

Additionally, the mean errors of location estimation range from 0.257 to 0.316 meters, which are relatively small. In conclusion, the proposed detection based neural network demonstrates robust and reliable performance in accurately estimating the locations of different cars.

Fig. \ref{cdf}(b) shows the mean localization error as a function of distance. It demonstrates that while location estimates from images cannot provide accurate CSI, they can infer channel characteristics and are less affected by the distance between UEs and APs. %As shown in Fig. \ref{cdf}(b), the mean localization error remains relatively stable across various distance ranges, indicating that the image detection network maintains consistent accuracy even as the distance increases.
 This stability contrasts with traditional RF pilot based methods, where estimation accuracy degrades with increased distance. The results underscore the robustness of the proposed multimodal approach, especially when users are far from APs but within the camera's field of view.

\subsection{ Performance of User Association and Beamforming GNN}

 \begin{figure*}[tbp]
\centerline{\includegraphics[width=1 \linewidth]{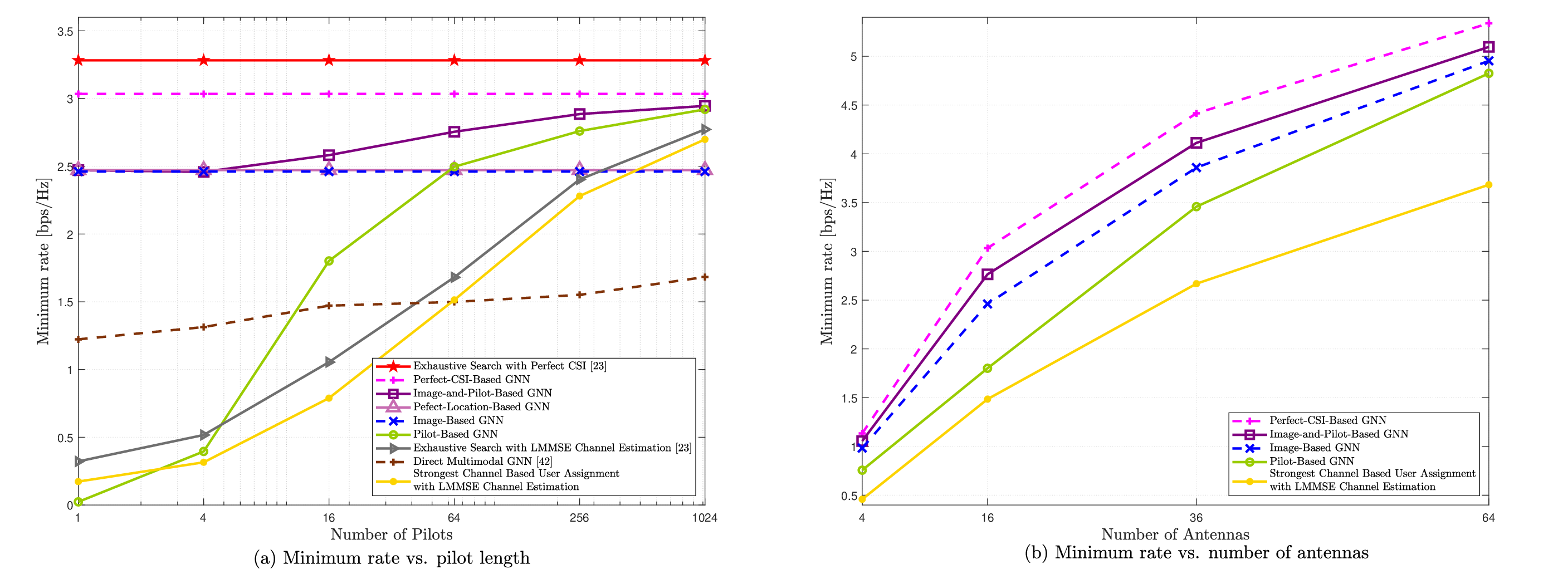}}
\caption{ Minimum rate performance of the proposed GNN {with $K=6$ and $L=4$. }}
\label{performance}
\vspace{-2pt}
\end{figure*}
We illustrate the impact of the number of uplink pilots on the downlink minimum rate performance for different methods in Fig. \ref{performance}(a). In the simulation, the number of antennas is set to 16.
All methods are evaluated on the same test dataset to ensure a fair comparison, and the performance is averaged over 200 samples to ensure the reliability of the results.

In Fig.~\ref{performance}(a), the exhaustive search algorithm with perfect CSI method consistently achieves the highest minimum rate of 3.28 bps/Hz across all pilot numbers. This method serves as an upper bound on the performance, since it utilizes perfect CSI to evaluate every possible user association configuration to find the most effective solution. The proposed GNN architecture achieves a performance with only about a $7.6\%$ reduction in the downlink minimum rate as compared with exhaustive search with perfect CSI. This indicates that the proposed GNN structure is capable of learning and approximating the exhaustive search based method effectively.

{Furthermore, the Image-Based GNN performs closely to the Perfect-Location-Based GNN method, which shows the efficiency of the proposed image based location detection neural network in providing accurate location estimates for system optimization.}

The pilot based GNN demonstrates a notable improvement in minimum rate performance as the number of pilots increases. From \eqref{eq:pilot2}, it is evident that the obtained CSI approaches the perfect CSI with an increasing number of pilots. Consequently, reliable channel information derived from pilots leads to improved overall system performance. 
The Image-Based GNN performs well in environments with limited pilot signals because it provides accurate location information independent of the number of pilot signals. {The minimum rate performance of the Direct Multimodal GNN is much lower than that of the Image-and-Pilot-Based GNN, which indicates that it is challenging for a single GNN to effectively learn from different data sources without separate feature extraction. }
The proposed multimodal GNN, which integrates image and pilot data, consistently outperforms the approach of using images alone or using pilots alone. The superior performance of the multimodal GNN indicates that the proposed GNN-based neural network effectively integrates features extracted from images and RF pilot signals and effectively utilizes the combined information for efficient system optimization.

Other traditional methods fall short as compared to the proposed GNN-based approaches, further demonstrating the effectiveness of the proposed method. In conclusion, the integration of visual data with pilot signals proves especially beneficial for improving the minimum rate performance compared with pilot-based benchmarks.

{Next, we illustrate the impact of the number of antennas on the minimum rate performance for different methods in Fig.~\ref{performance}(b), with the number of pilots set to 16. As the number of antennas increases, the minimum rate improves across all methods. The Perfect-CSI-Based GNN achieves the highest performance and serves as an upper bound. The proposed Image-and-Pilot-Based GNN achieves a competitive minimum rate as compared to the Perfect- CSI-Based GNN, which demonstrates the effectiveness of the proposed GNN across different numbers of antennas. Furthermore, the proposed Image-and-Pilot-Based GNN consistently outperforms the Pilot-Based GNN, which demonstrates the benefits of utilizing multimodal data for user association and beamforming optimization.}

{From Fig.~\ref{performance}(a) and Fig.~\ref{performance}(b), we observe that the proposed Image-and-Pilot-Based GNN consistently achieves a higher minimum rate than traditional and pilot-based methods, regardless of the number of pilots or antennas.  These results confirm the effectiveness of multimodal data fusion for optimizing user association and beamforming in wireless networks.}

\subsection{ Complexity Comparison}
\begin{table*}[ht]
\renewcommand{\arraystretch}{1.3} % Adjust row height
\centering
\caption{Average running time for solving one user association problem.}
\resizebox{0.87\textwidth}{!}{ % Resize the table to 90% of text width
\begin{tabular}{c c c c c c c}
\hline
\makecell{{Exhaustive Search} \\ {with Perfect CSI}} & 
\makecell{{Exhaustive Search} \\ {with LMMSE Estimation}} & 
\makecell{{Strongest-Channel-Based} \\ { Assignment} \\ {with LMMSE Estimation}} & 
\makecell{{Perfect-CSI-Based} \\ { GNN}} & 
\makecell{{Pilot-Based} \\ { GNN (GNN-P)}} & 
\makecell{{Image-Based} \\ {GNN (GNN-I)}} & 
\makecell{{Image-and-Pilot-Based} \\ { GNN (GNN-C)}} \rule[-3ex]{0pt}{6ex} \\ 
\hline
8.963 min & 
9.152 min & 
1.660 s & 
0.132 ms & 
0.136 ms & 
\makecell{{Detection:} 4.613 ms \\ {GNN-I:} 0.142 ms} & 
\makecell{{Detection:} 4.613 ms \\ {GNN-C:} 0.171 ms} 
\rule[-2ex]{0pt}{5ex} \\ 
\hline
\end{tabular}
}
\label{complexity}
\end{table*}

 To evaluate the computational complexity of the proposed methods, we compare the average running time required to solve a single instance of the problem. The results are averaged over 200 samples to ensure reliability.
All the methods are implemented on a computer equipped with a $\text{Intel}^{\circledR}$ $\text {Core}^{\mathrm{TM}}$ i7-12700 CPU (2.10GHz) and an NVIDIA GeForce RTX 4090 GPU. The results are summarized in Table \ref{complexity}. 

In Table \ref{complexity}, the exhaustive search methods exhibit the longest running time; the excessive running time makes them impractical for real-time applications.
In contrast, the strongest-channel-based assignment method offers a significantly faster solution; however, it sacrifices optimality, as observed in Fig.~\ref{performance}.
The GNN-based approaches demonstrate substantial improvements in efficiency, with all methods having running time lower than 5ms. The fast running time makes the proposed GNN suitable for real-time applications.
The proposed multimodal based GNN completes the location detection in 4.613ms and requires 0.171ms to process the features and to provide the final output. This method, while slightly slower than the image-only GNN and pilot-only GNN, offers a balanced trade-off between computational efficiency and the robustness of the solution.
Overall, the compelling advantages in both running time and performance of the proposed GNN-based approach make it well suited for practical deployment.

\subsection{ Generalization }

In this section, we evaluate the generalization capability of the proposed method.
We compare the performance of two GNN models: one trained with a fixed number of cars and another trained with varying number of cars. Both models are tested on a dataset with a fixed number of cars to ensure a consistent basis for comparison.
 \begin{figure*}[tbp]
\centerline{\includegraphics[width=0.9 \linewidth]{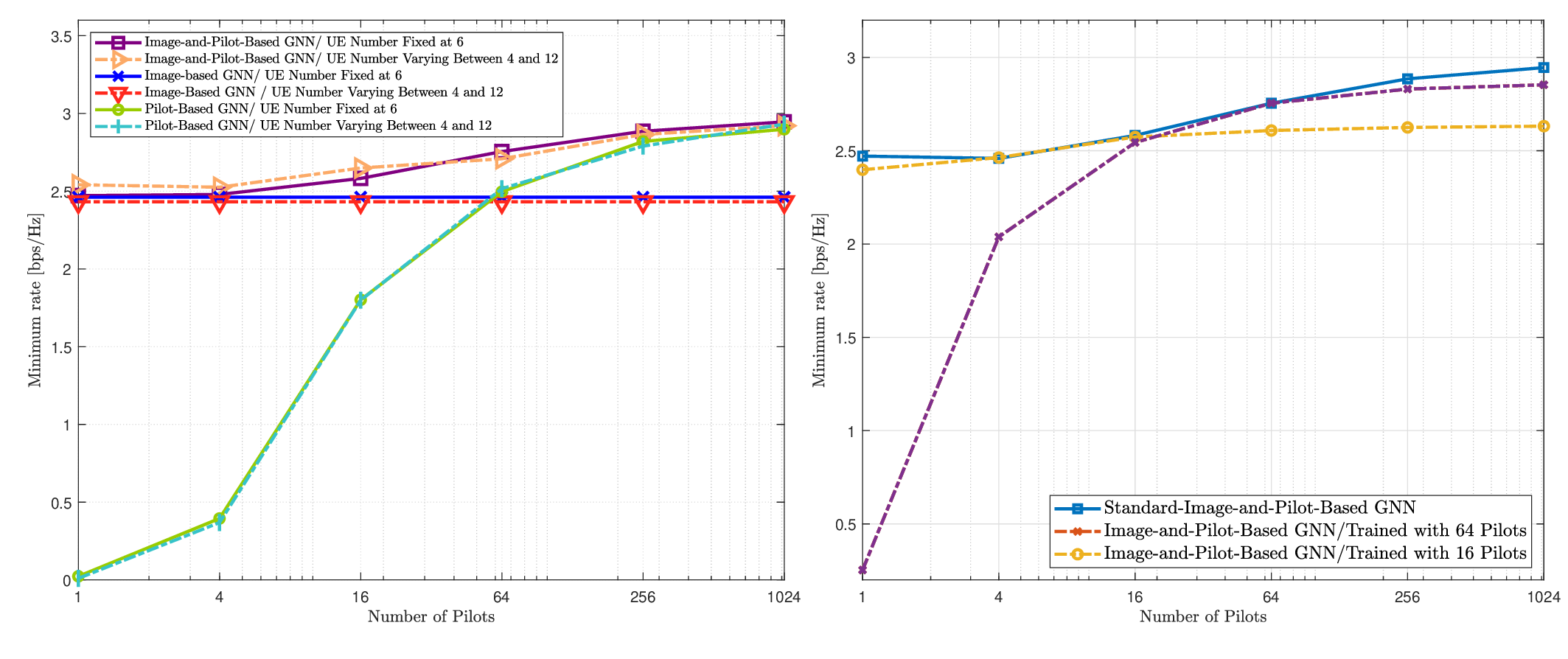}}
\caption{Generalization Performance of the Proposed GNN Models. a) Minimum rate performance with varying and fixed car (UE) numbers for training. b) Minimum rate performance of GNN models trained with different numbers of pilots.  }
\label{generalization}
\vspace{-2pt}
\end{figure*}
Fig.~\ref{generalization}(a) illustrates the minimum rate performance of the two GNN models as a function of the number of pilots. Although scenarios with varying numbers of cars introduce increased complexity and require more flexible decision-making, the proposed method maintains a similar minimum rate as the model trained with a fixed number of cars.
This result demonstrates the effectiveness of the proposed method in generalizing across scenarios with varying numbers of cars, which is essential for practical applications.
 
{
In Fig.~\ref{generalization}(b), we illustrates the minimum-rate performance of the GNN models trained with different numbers of pilots. The Standard-Image-and-Pilot-Based GNN is trained using a dataset where the number of pilots matches the number of pilots used during inference. 
The Image-and-Pilot-Based GNN ($N_u^\text{train}=\omega$) represents the GNN trained on a dataset with a fixed number of $\omega$ pilots, regardless of the number of pilots used during inference. To evaluate the generalization performance of the proposed GNN, we measure its minimum rate performance when the proposed GNN is trained with 16 pilots and 64 pilots, and then tested under different numbers of pilots.
When trained with 16 pilots, the GNN performs as well as the Standard-Image-and-Pilot-Based GNN when the number of pilots used for inference is low. However, when the number of pilots is significantly greater than 16, its performance fails to match the Standard-Image-and-Pilot-Based GNN. This is reasonable because a low number of pilots provides less reliable CSI, which limits the ability of the GNN to learn meaningful representations during the training process. In such cases, the GNN relies more on image-based information for decision-making.
When trained with 64 pilots, the proposed GNN method performs almost as well as the Standard-Image-and-Pilot-Based GNN when the number of pilots ranges from 16 to 1024. This strong performance across different pilot settings indicates that the GNN trained with 64 pilots has robust generalization capabilities.}

{
If the length of pilots can be provided as additional information, combining the results from Image-and-Pilot-Based GNN ($N_u^\text{train}=16$) and Image-and-Pilot-Based GNN ($N_u^\text{train}=64$) could achieve performance very close to that of the Standard-Image-and-Pilot-Based GNN across a pilot range from 1 to 1024. When the length of pilots is 16 or fewer, we choose the result from the GNN trained with $N_u^\text{train}=16$, and when it is greater than 16, we choose the result from the GNN trained with $N_u^\text{train}=64$.  The results demonstrate that with pilot information, two GNN models trained with a fixed number of pilots can be efficiently utilized to solve the user association and beamforming problems across a large range of pilot lengths.}

{
Using a dataset with varying number of cars for training the neural network leads to high variation in user distributions and makes it difficult for the neural network to realize effective optimization. Furthermore, achieving effective optimization on a dataset different from training requires high generalization capability of the trained neural network. Despite the challenges, the proposed method exhibits strong generalization properties. This is largely due to three key characteristics of the proposed GNN structure.
First, the structure of GNNs is designed to capture interactions between nodes (UEs) through graph structures rather than relying solely on memorizing data distributions. This property enables GNNs to naturally capture relational dependencies between UEs, regardless of the number of UEs in the system. Moreover, GNNs apply the same message-passing function to all nodes, which means GNNs are more data efficient. Lastly, the proposed method is less sensitive when the distribution of the data changes, because it extracts features separately using two different GNNs. This separate feature extraction ensures that changes in the distribution of one type of data do not interfere with feature extraction from the other type. These properties enable the proposed method to generalize effectively.
In conclusion, the results from Fig.~\ref{generalization} demonstrate strong generalization capability and effectiveness of the proposed GNN-based framework.}

 \begin{figure*}[tbp]
\centerline{\includegraphics[width=1 \linewidth]{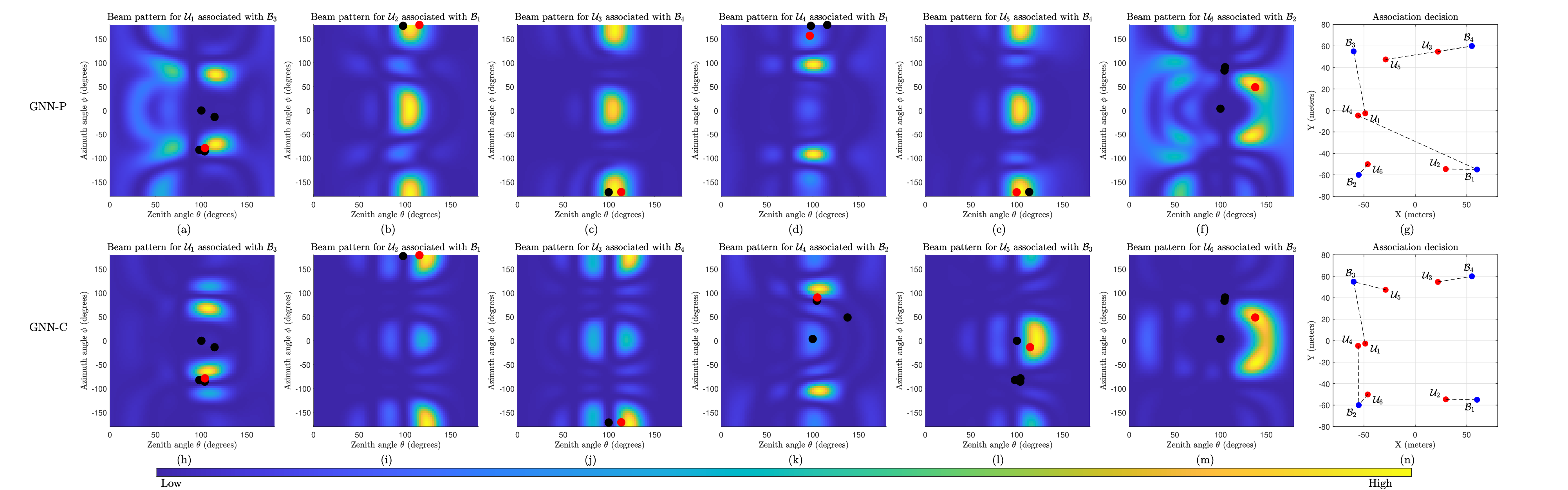}}
\caption{Array response of the APs and user association decisions for the cars obtained from GNN-P and GNN-C. The transmitted pilot length is set as 4.}
\label{interpretation}
\end{figure*}

\section{Interpretation of Solutions From GNN}

To interpret the solutions from the GNN, this section visually verifies that the designed beamformer maximizes the transmitted signal power at the angle of the dominant path from the target UE. Additionally, this section demonstrates that the user association configurations are designed based on the appropriate optimization criteria, rather than merely assigning users to their nearest APs. 
We use the array response function to qualitatively evaluate the correctness of the beamforming   vector learned by the proposed GNN. Let $\phi_{k}$ and $\theta_{k}$ represent the azimuth and zenith angles of arrival from $\mathcal{U}_k$ to the associated AP, respectively. Let $\mathbf{v}_k$ denote the learned beamforming vector for $\mathcal{U}_k$ from the associated AP. The array response for $\mathcal{U}_k$ is a function of the angles $\theta_{k}$ and $\phi_{k}$ defined by 
\begin{equation}
\label{eq:response}
   f_k(\theta_{k},\phi_{k}) = ||\mathbf{p}(\theta_{k},\phi_{k})^{\mathrm{H}} \mathbf{v}_k||^2_2,
\end{equation}
where
\begin{equation}
    [\mathbf{p}(\theta_{k},\phi_{k})]_m = e^{\frac{2 j \pi d_{\text{R}}}{\lambda_c} (v_1(m, M_c) \sin(\theta_{k}) \cos(\phi_{k}) +v_2(m, M_c) \cos(\theta_{k}))}.
\end{equation}

Fig.~\ref{interpretation}(a-f) presents the learned array responses for 6 UEs generated by GNN-P, while Fig.~\ref{interpretation}(h-m) shows the learned array responses generated by GNN-C. The color intensity represents the array response strength as defined in  (\ref{eq:response}). The red point represents the target user, while the black points represent other cars that are not the primary focus of the beamformer. The length of transmitted pilots, $N_u$, is set to 4. Fig.~\ref{performance} indicates that the minimum rate performance of GNN-C is better than that of GNN-P. From the array response pattern for $\mathcal{U}_2$, it is evident that the array response differs between GNN-P and GNN-C. Compared with GNN-P, the beamforming vectors generated by GNN-C are more accurately directed towards the target users. This difference indicates the proposed GNN-C's ability to effectively combine pilot and image information for a more precise and efficient beamforming vector design.

 %Typically, a car is assigned to the nearest AP, but if an AP is overloaded, users near multiple APs should be reassigned to less congested APs to balance the load.  The optimal user association rule is complex because it is based on many factors, such as the number of cars served by each AP and the distance between cars and APs.
Fig.~\ref{interpretation}(g) and  Fig.~\ref{interpretation}(n) illustrate the optimized user association decisions from the GNN-P and GNN-C. Each subplot shows the spatial distribution of APs and UEs. The UEs are represented by red dots, and the APs are represented by blue dots. If the dashed line connects a UE and an AP, it indicates that the UE is associated with that AP based on the optimized user association decisions.
To verify the effectiveness of the user association decisions, we compare the user association decision from the propose method with the simplest nearest-AP association rule.
Observing from Fig.~\ref{interpretation}(g), the UEs are not optimally associated with the APs when the number of pilot signals is set to 4. The reason is that the insufficient number of pilots leads to inaccurate CSI, which affects the optimization performance.   In contrast, as shown in Fig.~\ref{interpretation}(n), GNN-C demonstrates a more effective user association pattern by leveraging additional location information from the images. In Fig. \ref{interpretation}(n), most UEs are associated with the nearest AP. However, the user association decision for $\mathcal{U}_1$ does not simply follow the nearest AP association rule. As shown in Fig.~\ref{interpretation}, the Cartesian coordinates of $\mathcal{U}_1$ are [-48.6304, -2.6822, 0.4]. The distance between $\mathcal{U}_1$ and $\mathcal{B}_2$ is 59.49, and the distance between $\mathcal{U}_1$ and $\mathcal{B}_3$ is 60.58. Despite $\mathcal{U}_1$ being closer to $\mathcal{B}_2$, it is associated with $\mathcal{B}_3$ instead.
Under the max-min objective function, this user association configuration from GNN-C can be explained. Since $\mathcal{B}_2$ is already serving 2 users, associating $\mathcal{U}_1$ with $\mathcal{B}_2$ would likely result in a lower data rate as compared to associating it with $\mathcal{B}_3$, whose load is lighter. The small distance difference between $\mathcal{U}_1$ and the two APs is negligible when compared to the benefit of better load balancing.  
Comparing the performance of GNN-C and GNN-P, the proposed method, which integrates pilot and image data for system optimization, provides better system configurations than methods that rely solely on pilots.

%Since each AP has the same power limit, distributing the users more evenly across the available APs ensures that no single AP becomes a bottleneck.
\section{conclusion}
This paper introduces a learning-based approach that integrates visual images with RF pilots to solve the joint user association and beamforming optimization problem in a downlink wireless cellular network. The incorporation of images enhances the channel awareness of wireless communication systems and ensures robust system performance without relying on extensive pilot transmissions for channel estimation.
Specifically, we first propose a detection neural network to estimate the user locations from images. Then, we introduce a GNN-based neural network that effectively maps the location information derived from images and RF pilots to the optimized system configurations. %Additionally, a modified STE is employed to handle the non-differentiability of binary user associations during backpropagation. The modified STE ensures global optimization in decision-making by enabling continuous updates to both user associations and beamforming vectors, preventing premature exclusion of user-AP pairs throughout the training process.
Simulation results demonstrate that integrating visual data with RF signals provides superior optimization performance compared with methods relying solely on RF signals. The strong generalization capabilities and low computational complexity make the proposed GNNs suitable for real-time applications in dynamically varying user scenarios. 

\bibliographystyle{IEEEtran}
\bibliography{IEEEabrv,reference}

\begin{IEEEbiography}[{\includegraphics[width=1in,height=1.25in,clip,keepaspectratio]{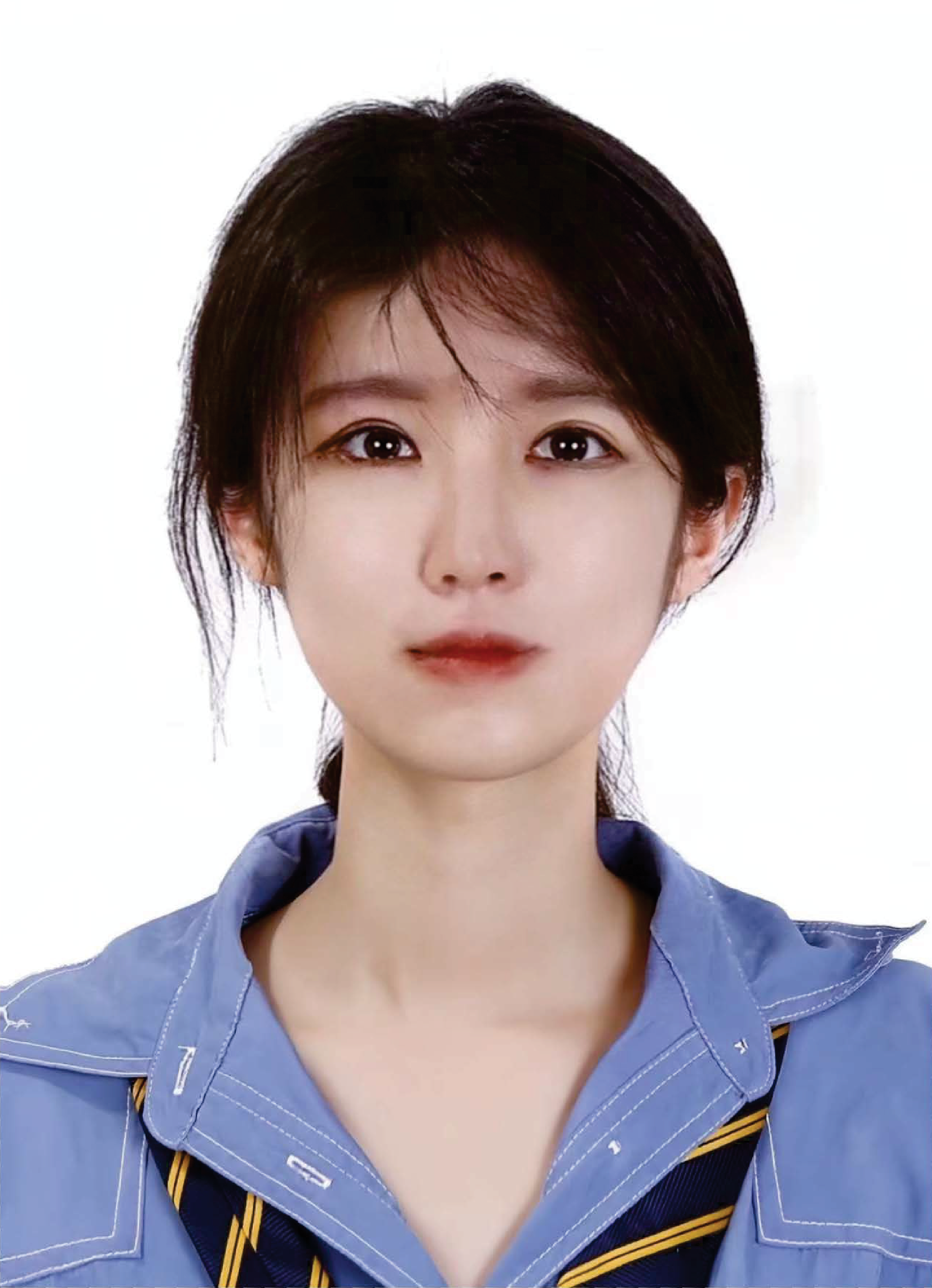}}]{Yinghan Li}
(Graduate Student Member, IEEE) received the B.S. and M.S. degrees in electronic and information engineering from Beihang University, Beijing, China, in 2018 and 2021, respectively. She is currently pursuing the Ph.D. degree in electrical and computer engineering at the University of Toronto, Toronto, ON, Canada. Her research interests include multimodal learning, wireless communications, and optimization.
\end{IEEEbiography}

\begin{IEEEbiography}[{\includegraphics[width=1in,height=1.25in,clip,keepaspectratio]{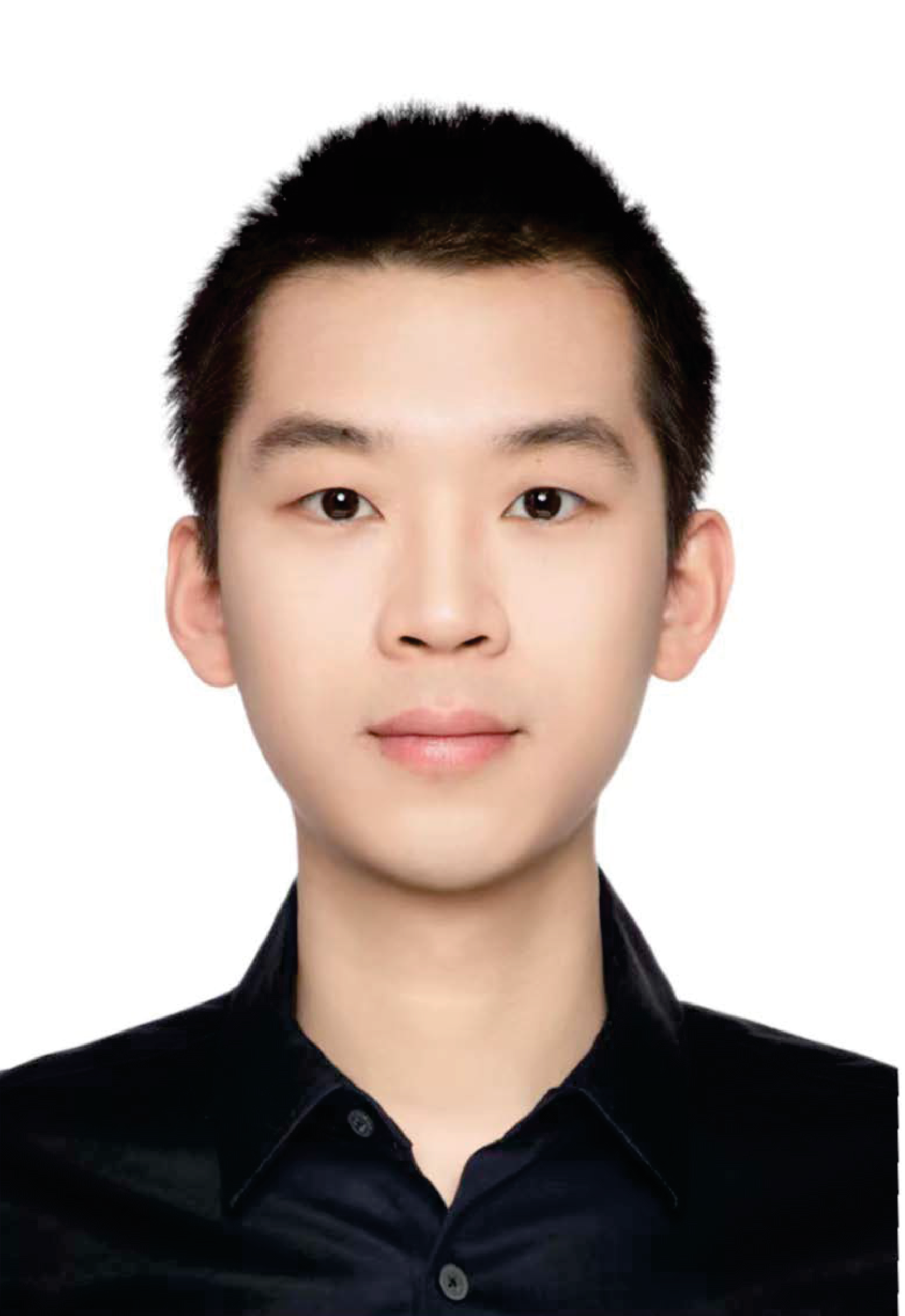}}]{Yiming Liu}(Graduate Student Member, IEEE) received the  M.S. degree in information and communication engineering from Tongji University, Shanghai, China, in 2022. He is currently pursuing the Ph.D. degree in electrical and computer engineering with the University of Toronto, Toronto, Ontario, Canada. His research interests include communications and sensing, signal processing, information theory, and machine learning. He was the recipient of the China National Scholarship in 2018 and 2021, the Shanghai Outstanding Graduate Award in 2022, the HKUST Redbird Ph.D. Award in 2022, and the Connaught International Scholarship in 2023. He was also recognized as an Exemplary Reviewer for \textsc{IEEE Wireless Communications Letters} in 2022.
\end{IEEEbiography}
\begin{IEEEbiography}[{\includegraphics[width=1in,height=1.25in,clip,keepaspectratio]{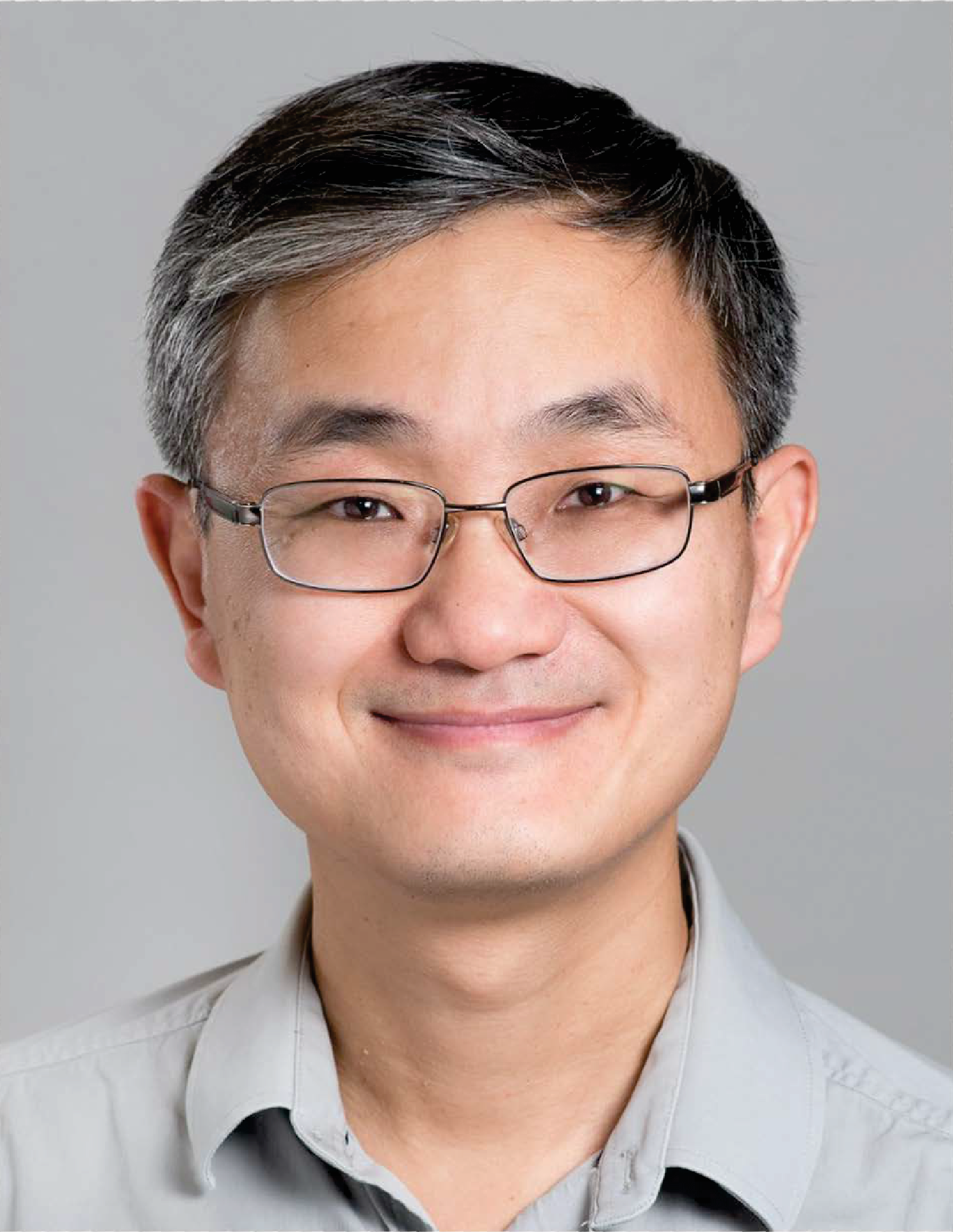}}]{Wei Yu}
(Fellow, IEEE) received the B.A.Sc. degree in computer engineering and mathematics from the University of Waterloo, Waterloo, ON, Canada, and the M.S. and Ph.D. degrees in electrical engineering from Stanford University, Stanford, CA, USA. He is currently a Professor with the Electrical and Computer Engineering Department, University of Toronto, Toronto, ON, Canada, where he holds a Canada Research Chair (Tier 1) in Information Theory and Wireless Communications. He is a Fellow of the Canadian Academy of Engineering and a member of the College of New Scholars, Artists, and Scientists of the Royal Society of Canada. He received the Steacie Memorial Fellowship in 2015, the IEEE Marconi Prize Paper Award in Wireless Communications in 2019, the IEEE Communications Society Award for Advances in Communication in 2019, the IEEE Signal Processing Society Best Paper Award in 2008, 2017, and 2021, the Journal of Communications and Networks Best Paper Award in 2017, the IEEE Communications Society Best Tutorial Paper Award in 2015, the IEEE Communications Society and Information Theory Society Joint Paper Award in 2024, and the R. A. Fessenden Award from IEEE Canada in 2024. He served as the President of the IEEE Information Theory Society in 2021, and as the Chair of the Signal Processing for Communications and Networking Technical Committee of the IEEE Signal Processing Society from 2017 to 2018. He served as an Area Editor for IEEE Transactions on Wireless Communications, an Associate Editor for \textsc{IEEE Transactions on Information Theory}, and an Editor for \textsc{IEEE Transactions on Communications} and \textsc{IEEE Transactions on Wireless Communications}. He was an IEEE Communications Society Distinguished Lecturer in 2015-16. He is a Clarivate Highly Cited Researcher.
\end{IEEEbiography}

\vfill

\end{document}